\documentclass[onecolumn]{autart}    

\usepackage{graphicx}          

\usepackage[sort,compress]{cite}

\usepackage{amssymb,amscd,amssymb,amsmath,mathrsfs}
\usepackage{epstopdf}
\usepackage{balance}
\usepackage[noend]{algorithm} 
\usepackage{algorithmicx} 
\usepackage[noend]{algpseudocode}
\usepackage{multirow} 
\usepackage{amsmath}
\usepackage{xcolor}
\usepackage{subfigure}
\usepackage{setspace}
\usepackage{enumerate}
\usepackage{lscape}
\usepackage{framed}
\usepackage[noend]{algpseudocode}
\usepackage{algorithmicx,algorithm}
\floatname{algorithm}{Algorithm}

\makeatletter

\newcommand{\Rmnum}[1]{\expandafter\@slowromancap\romannumeral #1@}
\makeatother

\emergencystretch=\maxdimen
\theoremstyle{definition}
\newtheorem{example}{Example}

\theoremstyle{remark}
\newtheorem{remark}{Remark}

\begin{document}

\begin{frontmatter}

\title{Quantitatively Nonblocking Supervisory Control of Discrete-Event Systems}


\author[Zhang]{Renyuan Zhang}\ead{ryzhang@nwpu.edu.cn},    
\author[Zhang]{Jiahao Wang},
\author[Zhang]{Zenghui Wang},
\author[Cai]{Kai Cai}\ead{cai@omu.ac.jp}  

\address[Zhang]{Northwestern Polytechnical University, Xi'an, China}  
\address[Cai]{Osaka Metropolitan University, Osaka, Japan}        


\begin{abstract}                         
In this paper, we propose a new property of {\em quantitative nonblockingness} of automata for partitions on the marker state sets of the automata.
This property {\em quantifies} the standard nonblocking property
by capturing the practical requirement that every subset (in the partition) of marker states (representing the same marking information) can be reachable within a prescribed number
of steps from any reachable state and following any trajectory of the system.
Accordingly, we formulate a new problem of quantitatively nonblocking supervisory control, and characterize its solvability in terms of a new concept of quantitative language completability.
It is proven that there exists the unique supremal quantitatively completable sublanguage of a given language, and we develop an effective algorithm to compute the supremal sublanguage. Finally, combining with the algorithm of computing the supremal controllable sublanguage, we design an algorithm to compute the maximally permissive solution to the formulated quantitatively nonblocking supervisory control problems.
\end{abstract}

\end{frontmatter}

\setlength{\abovedisplayskip}{0.2cm}
\setlength{\belowdisplayskip}{0.2cm}

\section{Introduction}
In standard supervisory control of discrete-event systems (DES) \cite{RamWon87,WonRam87,RamWon89,Wonham16a,CaiWon20,WonCaiRud18,CasLaf08}, and other extensions and applications on nonblocking supervisory control, e.g. \cite{BalemiEt:1993, BrandinCharbonnier:1994, KumarShayman1994, FabianKumar:1997, Malik:2003, MaWonham:2005, QueirozEt:2005}, the plant to be controlled is  modeled by finite-state automata and {\em marker states} are used to represent `desired states'. A desired state can be a goal location, a start/home configuration, or a task completion \cite{Elienberg:1974, Wonham16a}.  Besides enforcing all imposed control specifications, a {\em nonblocking supervisor} ensures that every system trajectory can reach a marker state (in a finite number of steps). As a result, the system under supervision may always be able to reach a goal, return home, or complete a task.

While the nonblocking property is important, it only {\em qualitatively} guarantees finite reachability of marker states. There is no given bound on the number of steps for reaching marker states, so it can take an arbitrarily large (though finite) number of steps before a marker state is reached. Consequently, this qualitatively nonblocking property might not be sufficient for many practical purposes, especially when there are prescribed bounds for reaching desired states. For example, a production cell \cite{FengCW09} may be required not only to complete a task (e.g. transporting/processing a batch of workpieces) but also to do so within a prescribed number of operations; a warehouse AGV \cite{Gagliardi:2012} is typically expected not only to return to a self-charging area but to do so periodically with a predetermined period (described by a number of events, each representing a movement from one area to the next); a communication protocol \cite{Milner89} is required not only to complete sending of a message and receiving of an acknowledgement, but also to do so in a bounded number of sending/{receiving} operations. In Section 2 below, we will present a detailed motivation example.

{
With the above motivation, we propose a {\em quantitatively} nonblocking property of an automaton to capture the practical
requirement that for a given partition on the marker state set, each cell of the partition (representing a type of task) must be reached within a prescribed number $N_i$ of steps from any reachable state and following any string.
Roughly speaking, we measure the `maximal distance' between reachable states and the specified subset of marker states, and this is done by counting the number of events in {\it every} string leading a reachable state to one of marker states in the specified subset.  More specifically, assume that the marker state set of the plant is partitioned according to $\{Q_{m,i}|i \in \mathcal{I}\}$ ($\mathcal{I}$ an index set),
and let $N_i$ be a finite positive integer which denotes the required number of steps to reach marker states in $Q_{m,i}$. We define a {\em quantitatively nonblocking property} (with respect to $\{(Q_{m,i}, N_i) | i \in \mathcal{I}\}$) of an automaton that from every reachable state, all the strings that lead the state to a marker state in $Q_{m,i}$ have lengths smaller than or equal to $N_i$ for all $i \in \mathcal{I}$. That is, in the worst case, for every marker state subset $Q_{m,i}$, every reachable state can reach one of marker states in $Q_{m,i}$ in no more than $N_i$ steps following any string.
If each marker state subset $Q_{m,i}$ represents the completion of a type of task, this quantitatively nonblocking property requires the automaton to be able to complete all types of tasks of the plant in at worst $N_i$ steps following all possible trajectories. Hence, we treat all the marker states in a subset $Q_{m,i}$ to be the same, where marker states in different subsets $Q_{m,i}$, $Q_{m,j}$ ($i \neq j$) are different.

Moreover, we formulate a new {\em quantitatively nonblocking supervisory control problem} ({\em QNSCP}) by requiring a supervisory control solution to be implementable by a quantitatively nonblocking automaton. To solve this problem, we present a necessary and sufficient condition by identifying a new language property called {\em quantitative completablility}. The latter roughly means that in the worst case, for every sublanguage $K_i$ ($i\in \mathcal{I}$) (defined according to a particular type of task corresponding to $Q_{m,i}$) of a given language $K$, every string in the closure of $K$ can be extended to a string in the sublanguage $K_i$ in no more than $N_i$ steps. Further we show that this language quantitative completability is closed under set unions, and together with language controllability which is also closed under unions, a maximally permissive solution exists for the newly formulated QNSCP. Finally we design polynomial algorithms for the computation of such an optimal solution.
}

We contrast our newly proposed concepts with other similar ones in the literature.
First, several other extensions of the standard nonblocking property have been studied. Multitasking supervisory control \cite{QueirozEt:2005} requires that every task must be completed, and for this the concept of strong
nonblockingness is proposed; this concept is similar to our concept of quantitative nonblockingness, but
does not consider the bound of transition steps on completing the tasks. We also explain in Section 3 that the method in \cite{QueirozEt:2005} cannot be directly adopted to solve QNSCP in our paper. \cite{MalikLudec:2008} introduces a concept of generalized nonblocking, which defines the coreachability between reachable states to subset of marker states representing particular properties.
\cite{DietrichEt:2002} proposes a stronger concept of nonblockingness by restricting
that specified marker states can be arrived by controllable paths consisting of a subset of controllable events.
\cite{WareMalik:2014, WareMalik:2014Conf} proposes another generalization of the nonblocking property by introducing the concept of progressive events and only these events can be used in strings towards task completion. However, all the concepts/properties mentioned above do not consider the requirement on the number of steps, which is in contrast with this work.
We also note that in \cite{WareMalik:2014, WareMalik:2014Conf, DietrichEt:2002}, it is assumed that the supervisor
can use special controllable events \cite{DietrichEt:2002} or progressive events \cite{WareMalik:2014, WareMalik:2014Conf}
to find a suitable path to steer the system to marker states. By contrast, this work does not make
such an assumption and only considers the most basic setup in which a supervisor is limited to enabling/disabling
controllable events. In this setup we study the problem of ensuring that all paths from all reachable states to
marker states are bounded by a given number $N$, which is in fact a weaker problem in this setup than finding a single path to marker states (see Section 2.2)

Second, the concept of $N$-step coreachability in quantitative nonblockingness is similar to bounded liveness in model checking \cite{BerardEt:2001}, which describes
the property that ``desired situations" must occur with a maximal delay. For bounded liveness, there are various
algorithms \cite{ChatterjeeEt:2021, ChatterjeeEt:2014, BiereEt:2003} to find strategies satisfying bounded liveness. However,
unlike our considered supervisory control problems, uncontrollable events and maximal permissiveness of strategies
are not considered.

{
This paper also distinguishes from its conference precursor \cite{ZhangWangCai:2021} by extending the concept of quantitative nonblockingness to a more general case,
where the marker state set of the plant is divided according to a given partition, and the requirement on the steps of reaching every maker state subset can be different.
Also, this paper provides all the proofs of formal results that are not given in \cite{ZhangWangCai:2021}.
}

This paper is organized as follows. Section 2 provides preliminaries and a motivating example.
Section 3 introduces the new concepts of quantitative nonblockingness of automata and quantitative completability of languages, and formulates the problem of QNSCP.
Section 4 presents a necessary and sufficient condition for solvability of QNSCP, and develop algorithms
to compute the supremal quantitatively completable sublanguage of a given language. Section 5 presents an effective solution to the QNSCP,
and finally Section 6 states our conclusion and future work.

\section{Preliminaries and Motivating Example}

In this section, we review the standard nonblocking supervisory control theory of DES \cite{RamWon87,WonRam87,Wonham16a} and present a motivating example for our work.

\subsection{Nonblocking Supervisory Control of DES}

A DES plant is modeled by a {\em generator} (or {\em automaton})\footnote{In the following we will use ``generator" and ``automaton" interchangeably. In this paper, the generators/automata representing the plant models and languages are assumed to be deterministic.} \cite{Wonham16a}
\begin{align} \label{eq:plant}
{\bf G} = (Q,\Sigma, \delta, q_0, Q_m)
\end{align}
where $Q$ is the finite state set; $q_0 \in Q$ is the initial state;
$Q_m\subseteq Q$ is the subset of marker states; $\Sigma$ is the finite event set;
$\delta: Q\times \Sigma\rightarrow Q$ is the (partial) state transition function.
Let $\Sigma^*$ be the set of all finite-length strings of events in $\Sigma$, including the empty string $\epsilon$.
In the usual way, $\delta$ is extended to $\delta:Q\times\Sigma^*\rightarrow Q$,
and we write $\delta(q,s)!$ to mean that $\delta(q,s)$ is defined.
The {\it closed behavior} of $\bf G$ is the language
$L({\bf G}) = \{s\in \Sigma^*|\delta(q_0,s)!\}\subseteq \Sigma^*$ and the {\it marked behavior} is
$L_m({\bf G}) = \{s\in L({\bf G})|\delta(q_0,s)\in Q_m\}$ $\subseteq L({\bf G})$.
A string $s_1$ is a {\it prefix} of a string $s$, written $s_1\leq
s$, if there exists $s_2$ such that $s_1s_2 = s$. For a string $s \in \Sigma^*$, write $\bar{s} := \{s_1 \in \Sigma^* \mid s_1 \leq s\}$ for the set of all prefixes of $s$. Note that $\epsilon$ and $s$ are members of $\bar{s}$.
For a { (regular)} language $K \subseteq L_m({\bf G})$,\footnote{All the languages discussed in this paper are assumed to be regular and thus
can be represented by finite state generators/automata \cite{Hopcroft14,Wonham16a}.}
the {\it (prefix) closure} of $K$ is $\overline{K} := \{s_1 \in
\Sigma^*|(\exists s\in K)~s_1\leq s\}$. We say that $K$ is {\em closed} if $K = \overline{K}$.

For a generator $\bf G$ as in (\ref{eq:plant}), a state $q \in Q$ is \emph{reachable} if there is a string $s \in L({\bf G})$ such that $q = \delta(q_0, s)$; state $q \in Q$ is \emph{coreachable} \cite{Elienberg:1974, Wonham16a}
 if there is a string $s \in \Sigma^*$ such that $\delta(q, s)!$ and $\delta(q, s) \in Q_m$.
We say that ${\bf G}$ is \emph{nonblocking} if every reachable state in ${\bf G}$ is coreachable.
In fact ${\bf G}$ is \emph{nonblocking} if and only if
$\overline{L_m({\bf G})} = L({\bf G})$  \cite{Wonham16a}.

For two generators ${\bf G}_i = (Q_i,\Sigma,\delta_i, q_{0,i}, Q_{m,i})$, $i = 1, 2$, their product generator is defined
as ${\bf G}_1 \times {\bf G}_2 = $ $(Q,\Sigma,\delta, q_{0}, Q_{m})$, where $Q = Q_1\times Q_2$, $\delta = \delta_1\times \delta_2$, $q_0 = (q_{0,1},q_{0,2})$,
and $Q_{m} = Q_{m,1}\times Q_{m,2}$, with $(\delta_2\times\delta_2)((q_1,q_2),\sigma):= (\delta_1(q_1,\sigma),\delta(q_2,\sigma))$ \cite{Wonham16a}.

For the control purpose, the event set $\Sigma$ is partitioned into
$\Sigma_c$ (the subset of {\em controllable} events) and $\Sigma_{uc}$ (the subset of
{\em uncontrollable} events), i.e. $\Sigma = \Sigma_c\dot\cup\Sigma_{uc}$.
A {\it supervisory control} for $\bf G$ is any map $V:L({\bf G})\rightarrow \Gamma$,
where $\Gamma := \{\gamma \subseteq \Sigma \mid \gamma \supseteq \Sigma_{uc}\}$. Then the
{\em closed-loop system} is denoted by $V/{\bf G}$, with closed behavior $L(V/{\bf G})$ defined as:
(i) $\epsilon \in L(V/{\bf G})$; (ii) $s \in L(V/{\bf G}) \ \&\ \sigma \in V(s) \ \&\ s\sigma \in L({\bf G}) \Rightarrow ~s\sigma \in L(V/{\bf G})$; (iii) no other strings belong to $L(V/{\bf G})$.
On the other hand, for any sublanguage $K \subseteq L_m({\bf G})$, the closed-loop system's marked behavior $L_m(V/{\bf G})$ is given by\footnote{With this definition of $L_m(V/{\bf G})$, the supervisory control $V$ is also known as a {\em marking supervisory control} for $(K, {\bf G})$ \cite{Wonham16a}.}
$L_m(V/{\bf G}) := K \cap L(V/{\bf G})$.
The closed behavior $L(V/{\bf G})$ represents the strings generated by the plant ${\bf G}$ under the control of $V$,
while the marked behavior $L_m(V/{\bf G})$ represents the strings that have some special significance, for instance
representing `task completion'. We say that $V$ is {\it nonblocking} if
\[\overline{L_m(V/{\bf G})} = L(V/{\bf G}).\]

\vspace{-0.3cm}

A language $K\subseteq L_m({\bf G})$ is {\em controllable} (w.r.t. ${\bf G}$ and $\Sigma_{uc}$) if $\overline{K}\Sigma_{uc}\cap L({\bf G}) \subseteq \overline{K}$. The following is a central result of nonblocking supervisory control theory  \cite{WonRam87,Wonham16a}.

\begin{thm} \label{thm:sct}
Let $K\subseteq L_m({\bf G})$, $K \neq \emptyset$. There exists a nonblocking (marking) supervisory control $V$ (for $(K, {\bf G})$) such that $L_m(V/{\bf G}) = K$ if and only if $K$ is controllable. Moreover, if such a nonblocking supervisory control $V$ exists, then it may be implemented by a nonblocking generator ${\bf S}$, i.e. $L_m({\bf S}) = L_m(V/{\bf G})$. \hfill $\diamond$
\end{thm}

Further, the property of language controllability is closed
under set union. Hence for any language $K \subseteq L_m({\bf G})$ (whether or not controllable), the set \[\mathcal{C}(K) = \{K' \subseteq K \mid \overline{K'}\Sigma_{uc}\cap L({\bf G}) \subseteq \overline{K'} \}\] contains a unique supremal element denoted by $\sup \mathcal{C}(K)$ \cite{WonRam87,Wonham16a}.  Whenever $\sup \mathcal{C}(K)$ is nonempty, by Theorem~\ref{thm:sct} there exists a nonblocking supervisory control $V_{\sup}$ that satisfies $L_m(V_{\sup}/{\bf G}) = \sup \mathcal{C}(K)$ and may be implemented by a nonblocking generator ${\bf SUP}$ with \[L_m({\bf SUP}) = L_m(V_{\sup}/{\bf G}).\]

\subsection{Motivating Example}

Nonblockingness of supervisory control $V$ describes a qualitative requirement that every string
generated by the closed-loop system $V/{\bf G}$ can be completed to
a marked string in finite but indefinite steps. However, in many real-world applications, it is often required that a task be completed in a prescribed, bounded number of steps from any system state and following any trajectory of the system. As an illustration, we present the following example.

\begin{figure}[!t]
\centering
    \includegraphics[scale = 0.34]{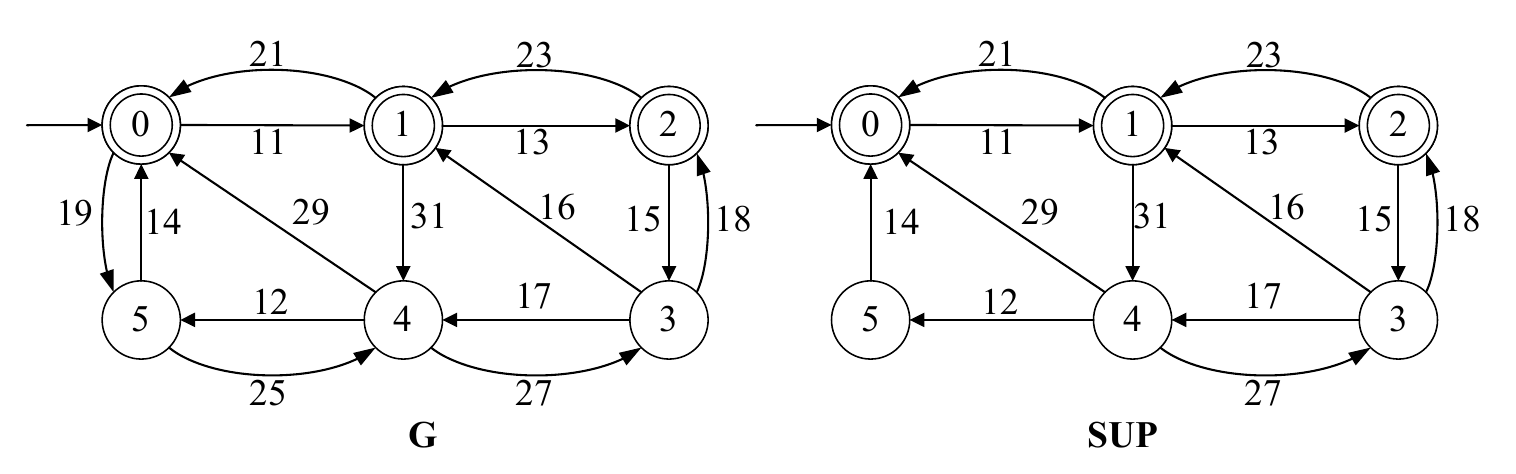}
\caption{Transition graphs of $\bf G$ and $\bf SUP$ } \label{fig:GSUP}
\vspace{-0.2cm}
\end{figure}

\begin{example}
Consider an autonomous vehicle for package collecting and delivery in a region.
The vehicle can move in six zones numbered 0--5, following the routes displayed on the top of Fig.~\ref{fig:GSUP}.
Zones 0 is the charging area for the vehicle to charge its battery.
Zones 1 and 2 are two service areas for customers where the customers can both receive packages from the vehicle
and call the vehicle to come to collect packages to be sent. Zones 3, 4 and 5 are the storage areas for incoming and outgoing
packages. Namely, the task of the vehicle is to send packages in the storage areas (zones 3, 4 and 5) to the service
areas (zones 1 and 2), and collect packages from the service areas
and store them into the storage areas. Also, the vehicle must be able to make a self-charging when it is running out of battery.

We model the movement of the autonomous vehicle by a generator $\bf G$ with transition graph displayed on the left of Fig.~\ref{fig:GSUP}.
States~0, 1 and 2 are chosen to be marker states; state~0 represents vehicle being charged, while states~1, 2
represent the completion of a received task.
We assume that the odd numbers represent controllable events and even numbers represent uncontrollable events.

First we consider an instance of standard nonblocking supervisory control. Suppose that due to road maintenance, the (directed) route
\vspace{-0.2cm}
\[\text{zone~0 $\rightarrow$ zone~5 $\rightarrow$ zone~4}\]
is not usable, {namely, the vehicle cannot move from zone 0 to zone 5, nor from zone 5 to zone 4.} This constraint is imposed as a specification.
To satisfy this specification, a nonblocking supervisory control can be synthesized \cite{WonRam87,Wonham16a}, and implemented by a nonblocking generator ${\bf SUP}$ as displayed on the {right side of Fig.~\ref{fig:GSUP}}.
This  {\bf SUP} disables event~19 at state~0 and event~25 at state~5.
Moreover, since {\bf SUP} is nonblocking, every {reachable} state can reach marker states 0, 1 and 2 in a finite number of steps.

Now consider two additional requirements that the customers need timely services:
\begin{enumerate}[{\rm (i)}]
 \item Every package sent to customers  must be delivered by the vehicle to either one of the two service areas (zone~1 or 2) within three steps (one step means the movement of vehicle from one zone to the next); and whenever a customer calls for package collection, the vehicle must reach either zone~1 or 2 within three steps no matter where the vehicle is and no matter which trajectory the vehicle follows.

 \item {The vehicle must be able to return to zone 0 for charging its battery within five steps.}
\end{enumerate}


The nonblocking supervisor $\bf SUP$ in Fig.~\ref{fig:GSUP} fulfills neither of the above additional requirements,
because if the vehicle is at zone~4, it is not guaranteed to return to zone 0 in five steps or to zone 1, 2 in three steps as it may move between zones 3 and 4 repreatedly.
Thus we need new concepts and methods that can quantify the number of steps of all possible paths from a reachable state to the specified (subsets of) marker states, and design new supervisors to satisfy the quantitative requirement on reaching marker states.

It needs to be stressed that we study this problem in the most basic setup of supervisory control: namely a supervisor
can only enable/disable controllable events. No further assumption on special events is made. In this basic
setup, finding one controllable path from a {reachable} state to a marker state is in fact a very strong requirement.
For example if the vehicle is at zone 3, a path of length one to reach zone 1 is ``16" and to reach zone 2 is ``18".
Neither path alone, however, is controllable (as events 16 and 18 are uncontrollable). In fact there does not
exist any controllable path from zone 3 to zone 1 or 2. This prompts us to consider not a single path but all
possible paths between {reachable states} and marker states.
\hfill $\diamond$
\end{example}



In the subsequent sections, we will formulate a problem of synthesizing {\em quantitatively} nonblocking supervisors, and provide an effective solution to the problem.


\section{Quantitatively Nonblocking Supervisory Control Problem Formulation}


We start by introducing a new concept that {\em quantifies} the nonblocking property of a generator.

Let ${\bf G} = (Q,\Sigma,\delta,q_0,Q_m)$ be a generator (modeling the plant to be controlled) as in (\ref{eq:plant}) and assume that ${\bf G}$ is nonblocking (i.e. every reachable state of ${\bf G}$ is also coreachable).
%
%
{
Bring in a partition $\mathcal{Q}_{\bf G}$ on the marker state set $Q_m$ as follows:\footnote{{It would be more general to consider a cover at the cost of introducing more technical assumptions. In this paper we choose to develop our theory based on partition for the sake of presentation clarity which is essential to convey the central idea of our work.}}
\begin{align} \label{eq:part}
\mathcal{Q}_{\bf G} := \{Q_{m,i} \subseteq Q_m | i \in \mathcal{I}\}.
\end{align}
Here $\mathcal{I}$ is an index set, $Q_{m,i} \neq \emptyset$ for each $i \in \mathcal{I}$, $Q_{m,i} \cap Q_{m,j} = \emptyset$ for all $i \neq j$, and $\bigcup \{Q_{m,i} | i \in \mathcal{I}\} $ $=  Q_m$.
This partition $\mathcal{Q}_{\bf G}$ represents a classification of different types of marker states. For example, the three marker states 0,1,2 in Example~1 can be classified into two types: $Q_{m,1} = \{1,2\}$ meaning completion of a package collecting/delivery task, whereas $Q_{m,2} = \{0\}$ meaning battery charging.


Fix $i \in \mathcal{I}$ and let  $q \in Q \setminus Q_{m,i}$ be an arbitrary state in $Q$ but not in $Q_{m,i}$.
We define the set of all strings that lead $q$ to $Q_{m,i}$ for the first time, namely
\begin{align*} 
C(q,Q_{m,i}):=\{s \in \Sigma^*| \delta(q,s)! ~\&~\delta(q,s) \in Q_{m,i}~\&\notag\\
     (\forall s' \in \overline{s}\setminus\{s\}) \delta(x,s') \notin Q_{m,i}\}.
\end{align*}
Note that $C(q,Q_{m,i})$ may be empty even though {\bf G} is nonblocking.
If $q \in Q_{m,i}$, we define $C(q,Q_{m,i}) := \{\epsilon\}$. 

Now associate $Q_{m,i}$ with a finite positive integer $N_i$, and consider an arbitrary state in $q \in Q$.
We say that state $q$ is {\em $N_i$-step coreachable} (wrt. $Q_{m,i}$) if
\begin{align*}
{\rm (i)} ~~ & C(q, Q_{m,i}) \neq \emptyset; \mbox{ and}\\
{\rm (ii)} ~~& (\forall s \in C(q,Q_{m,i})) ~|s| \leq N_i.
\end{align*}
Condition (i) requires that there exists a string $s \in \Sigma^*$ leading $q$ to a marker state in $Q_{m,i}$.
Condition (ii) means that all strings that lead $q$ to $Q_{m,i}$ for the first time have length at most $N_i$. 
Intuitively, condition~(ii) means that in the worst case, it takes $N_i$ steps from state $q$ to arrive a marker state in $Q_{m,i}$.
}

With $N_i$-step correachability, we introduce the new concept of quantitative nonblockingness of a generator.

\smallskip
\begin{defn} \label{defn:snnbg}
Let ${\bf G} = (Q, \Sigma, \delta, q_0, Q_m)$ be a generator, {$\mathcal{Q}_{\bf G} = \{Q_{m,i} | i \in \mathcal{I}\}$ a partition on $Q_m$ as defined in (\ref{eq:part}), and $N_i$ a positive integer associated with each $Q_{m,i} \in \mathcal{Q}_{\bf G}$.}
We say that ${\bf G}$ is \emph{quantitatively nonblocking} wrt. $\{({Q_{m,i}}, N_i)|i \in \mathcal{I}\}$ if for every $i \in \mathcal{I}$
and every reachable state $q \in Q$, $q$ is $N_i$-step coreachable (wrt. {$Q_{m,i}$}).
\end{defn}
\smallskip

In words, a quantitatively nonblocking generator requires that every state $q$ can reach every subset $Q_{m,i}$ of marker states within $N_i$ steps. In the special case where $\mathcal{Q}_{\bf G}$ is a partition with just one cell (i.e. $Q_{m,i} = Q_m$), all the marker states are treated the same and Definition~\ref{defn:snnbg} extends the standard concept of nonblockingness \cite{Wonham16a} by imposing a bound on the lengths of strings reaching $Q_m$. By the same reason (imposing bounds), Definition~\ref{defn:snnbg} is different from the concept of strong nonblockingness in \cite{QueirozEt:2005}.



Next we define the quantitatively nonblocking property of a supervisory control $V$. For this, we first introduce a new concept called {\em quantitative completability}.


Let $K \subseteq L_m({\bf G})$ be a sublanguage of $L_m({\bf G})$.
{For each marker state subset $Q_{m,i} \in \mathcal{Q}_{\bf G}$ define
\[L_{m,i}({\bf G}) := \{s \in L_m({\bf G}) | \delta(q_0, s) \in Q_{m,i}\}\]
i.e. $L_{m,i}({\bf G})$ represents the marked behavior of ${\bf G}$ wrt. $Q_{m,i}$. Then $K_i := K \cap L_{m,i}({\bf G})$, $i \in \mathcal{I}$.

For an arbitrary string $s \in \overline{K} \setminus K_i$, define the set of strings that lead $s$ to $K_i$ for the first time:
\begin{align} \label{eq:PKL}
M_{K,i}(s):=\{t \in \Sigma^* \mid st \in K_i 
(\forall t' \in \overline{t} \setminus\{t\}) st' \notin K_i \}.
\end{align}
If already $s \in K_i$, we define $M_{K,i}(s) :=\{\epsilon\}$. 
}

\smallskip
\begin{defn} \label{defn:sncomp}
Let ${\bf G} = (Q, \Sigma, \delta, q_0, Q_m)$ be a generator, $K  \subseteq L_m({\bf G})$ a sublanguage, {$\mathcal{Q}_{\bf G} = \{Q_{m,i} | i \in \mathcal{I}\}$ a partition on $Q_m$ as defined in (\ref{eq:part}), and $N_i$ a positive integer associated with each $Q_{m,i} \in \mathcal{Q}_{\bf G}$.}
For a fixed $i \in \mathcal{I}$, we say that $K$ is \emph{quantitatively completable} wrt. $({Q_{m,i}}, N_i)$ if
for all $s \in \overline{K}$,
\begin{align*}
{\rm (i)}&~ {M_{K,i}(s)} \neq \emptyset;\\
{\rm (ii)}&~ (\forall t \in {M_{K,i}(s)})~ |t| \leq N_i.
\end{align*}
Moreover if $K$ is quantitatively completable wrt. $({Q_{m,i}}, N_i)$ for all $i \in \mathcal{I}$, we say that $K$ is {\em quuantitatively completable} wrt. $\{({Q_{m,i}}, N_i)|i \in \mathcal{I}\}$.
\end{defn}
\smallskip

{If $K$ is quantitatively completable  wrt. $\{(Q_{m,i}, N_i)|i \in \mathcal{I}\}$, then for every $i \in \mathcal{I}$, every
string $s \in \overline{K} $ may be extended to a string in $K_i (= K \cap L_{m,i}({\bf G}))$  by strings of lengths at most $N_i$.}
We illustrate this definition by the following example.

\smallskip
\begin{example} [Continuing Example 1]
Consider the generator ${\bf G}$ in Example 1 (left of Fig.~\ref{fig:GSUP}), and let $K_1, K_2 \subseteq  L_m({\bf G})$ be sublanguages
as represented by generators ${\bf K}_1$ and ${\bf K}_2$ respectively (displayed in Fig.~\ref{fig:K12}). That is, $L_m({\bf K}_i) $ $ = K_i$, $i = 1, 2$.
\begin{figure}[!t]
\centering
    \includegraphics[scale = 0.36]{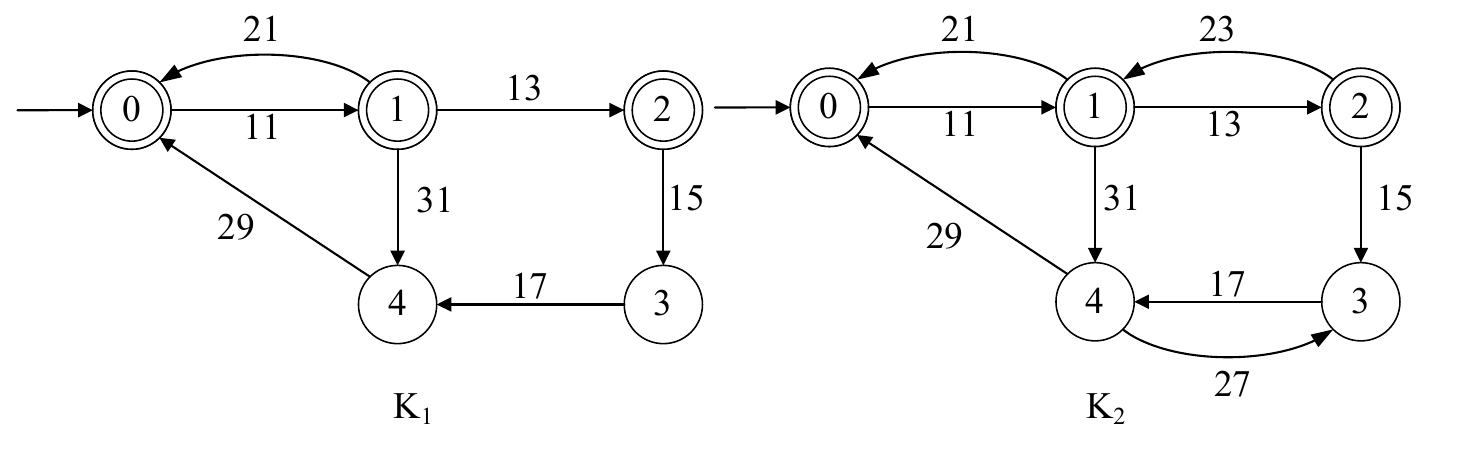}
\caption{Transition graphs of ${\bf K}_1$ and ${\bf K}_2$ } \label{fig:K12}
\vspace{-0.2cm}
\end{figure}

{Consider a 2-cell partition $\mathcal{Q}_{\bf G} = \{Q_{m,1}, Q_{m,2}\}$ on {\bf G}'s marker state set $\{0,1,2\}$, where $Q_{m,1} = \{1,2\}$ and $Q_{m,2} = \{0\}$. Also associate $N_1 = 3$ to $Q_{m,1}$ and $N_2 = 5$ to $Q_{m,2}$.
}
It is easily verified that $K_1$ is quantitatively completable wrt. {$\{(Q_{m,i}, N_i)| i = 1, 2\}$}: first for {$Q_{m,1} = \{1, 2\}$}, from states 0, 3, 4 all strings reaching marker state 1 or 2 are of lengths no more than three (e.g. state 3 reaches marker state 1 via string $17.29.11$ of length three);
second for {$Q_{m,2} = \{0\}$}, from states 1, 2, 3, 4 all strings reaching marker state 0 are of lengths no more than five (e.g. state 1 reaches marker state 0 via string $21$, string $35.29$, or string $13.15.17.29$, which have lengths 1, 2, 4 respectively).

However, $K_2$ is not quantitatively completable wrt. {$\{({Q_{m,i}}, N_i)|i =1,2\}$}. For {$Q_{m,1} = \{1, 2\}$}, from state 3 or 4, because of the loop between these two states, one may find a string (say 17.27.17.29.11 from state 3) that reaches $Q_{m,1}$ in more than three steps. Similarly for {$Q_{m,2} = \{0\}$}, the loop between states 1 and 2 allows a string longer than 5-step to reach state 0. Indeed, the existence of these loops makes $K_2$ is not quantitatively completable for any finite positive integers $N_1$ and $N_2$.
\hfill $\diamond$
\end{example}
\smallskip

{
The following result characterize the relation between quantitative completability of a language and quantitative nonblockingness of a generator.

\begin{prop} \label{prop:relation}
Let ${\bf G} = (Q, \Sigma, \delta, q_0, Q_m)$ be a nonblocking generator, $K \subseteq L_m({\bf G})$ a sublanguage, $\mathcal{Q}_{\bf G} = \{Q_{m,i} | i \in \mathcal{I}\}$ a partition on $Q_m$, and $N_i$ a positive integer associated with each $Q_{m,i} \in \mathcal{Q}_{\bf G}$.

{\rm (i)} If $K = L_m({\bf G})$ and ${\bf G}$ is quantitatively nonblocking wrt. $\{(Q_{m,i}, N_i)|i \in \mathcal{I}\}$, then $K$ is
quantitatively completable wrt. $\{(Q_{m,i}, N_i)|i \in \mathcal{I}\}$.

{\rm (ii)} If $K \subseteq L_m({\bf G})$ is quantitatively completable wrt. $\{(Q_{m,i}, N_i)|i \in \mathcal{I}\}$, then there exists a generator ${\bf K} = (X, \Sigma, \xi, x_0, X_m)$
such that $L_m({\bf K})=K$ and ${\bf K}$ is quantitatively nonblocking wrt. $\{(X_{m,i}, N_i)|i \in \mathcal{I}\}$,
where $X_{m,i} = \{x_m \in X_m| (\exists s \in \Sigma^*) \xi(x_0, s) = x_m \ \& \  \delta(q_0, s) \in Q_{m,i}\}$.
\end{prop}

{\it Proof:} First for part (i), when $K = L_m({\bf G})$, since $\bf G$ is nonblocking, we have for each $q \in Q$, there exists $s \in \Sigma^*$ such that $s \in L({\bf G}) = \overline{L_m({\bf G})} = \overline{K}$. Further, for each $t \in C(q,Q_{m,i})$ satisfying $\delta(q, t) \in Q_{m,i}$, we have $st \in L_{m,i}({\bf G}) = L_m({\bf G}) \cap L_{m,i}({\bf G}) = K \cap L_{m,i}({\bf G})$,
and for each $t' \in \overline{t}\setminus\{t\}$ satisfying $\delta(q, t) \notin Q_{m,i}$, we have $st \notin L_{m,i}({\bf G}) = L_m({\bf G}) \cap L_{m,i}({\bf G}) = K \cap L_{m,i}({\bf G})$.
Namely, when $K = L_m({\bf G})$, the definition of $C(q,Q_{m,i})$ is equivalent to that of $M_{K,i}(s)$. Thus if ${\bf G}$ is quantitatively nonblocking wrt. $\{(Q_{m,i}, N_i)|i \in \mathcal{I}\}$, then $K = L_m({\bf G})$ is
quantitatively completable wrt. $\{(Q_{m,i}, N_i)|i \in \mathcal{I}\}$.

For part (ii), let ${\bf K}'$ be a nonblocking generator representing ${K}$, i.e. $L_m({\bf K}') = K$ and $L({\bf K}') = \overline{K}$.
Then let ${\bf K} = (X, \Sigma, \xi, x_0, X_m)$ be the product generator of ${\bf K}'$ and {\bf G}. Since $K \subseteq L_m({\bf G})$, we have that $L_m({\bf K}) = K$, $L({\bf K}) = \overline{K}$, and {\bf K} is nonblocking. Moreover,  the marker state set $X_{m}$ is such that $X_{m} = \{x \in X | (\exists s \in \Sigma^*) \xi(x_0, s) = x \ \& \  \delta(q_0, s) \in Q_{m}\}$.

Now for the partition $\mathcal{Q}_{\bf G} = \{Q_{m,i} | i \in \mathcal{I}\}$ on $Q_m$, define $\mathcal{X}_{\bf K} = \{X_{m,i} | i \in \mathcal{I}\}$, where $X_{m,i} = \{x_m \in X_m | (\exists s \in \Sigma^*) \xi(x_0, s) = x_m \ \& \  \delta(q_0, s) \in Q_{m,i}\}$.
It is readily verified that $\mathcal{X}_{\bf K}$ is a partition on $X_m$.
Hence the definition of $M_{K,i}(s)$ is equivalent to that of $C(x,X_{m,i})$ wrt. {\bf K}. Therefore if $K$ is quantitatively completable wrt. $\{(Q_{m,i}, N_i)|i \in \mathcal{I}\}$, then ${\bf K}$ is quantitatively nonblocking wrt. $\{(X_{m,i}, N_i)|i \in \mathcal{I}\}$. The proof this is now complete. \hfill $\blacksquare$

According to Proposition~\ref{prop:relation}, for an arbitrary sublanguage $K \subseteq L_m({\bf G})$ that is quantitatively completable wrt. $\{(Q_{m,i}, N_i)|i \in \mathcal{I}\}$,
we may construct a quantitatively nonblocking (wrt. $\{(X_{m,i}, N_i)|i \in \mathcal{I}\}$) generator ${\bf K}$ representing $K$, i.e. $L_m({\bf K}) = K$. 
}

With the above quantitative completability of a language, we introduce the quantitatively nonblocking property of a supervisory control.
\smallskip
\begin{defn} \label{defn:snnbs}
Let ${\bf G} = (Q, \Sigma, \delta, q_0, Q_m)$ be a generator, $K  \subseteq L_m({\bf G})$ a sublanguage, {$\mathcal{Q}_{\bf G} = \{Q_{m,i} | i \in \mathcal{I}\}$ a partition on $Q_m$ as defined in (\ref{eq:part}), $N_i$ a positive integer associated with each $Q_{m,i} \in \mathcal{Q}_{\bf G}$.}
and $V : L({\bf G}) \rightarrow \Gamma$ a (marking) supervisory control (for $(K,{\bf G})$).
We say that $V$ is \emph{quantitatively nonblocking} wrt. {$\{(Q_{m,i}, N_i)|i \in \mathcal{I}\}$} if
\begin{align*}
{\rm (i)} ~~ &V \text{ is nonblocking}; \mbox{ and}\\
{\rm (ii)} ~~&L_m(V/{\bf G}) (=K \cap L(V/{\bf G}))~\text{is quantitatively} \\  &\text{completable wrt. {$\{(Q_{m,i}, N_i)|i \in \mathcal{I}\}$}.}
\end{align*}
\end{defn}
\smallskip

In words, quantitative nonblockingness of a supervisory control $V$ requires not only $V$ being nonblocking (in the standard sense), but also the marked behavior $L_m(V/{\bf G})$ of
the closed-loop system $V/{\bf G}$ being quantitatively completable. According to Proposition~\ref{prop:relation},
$L_m(V/{\bf G})$ can be represented by a quantitatively nonblocking generator.

We are ready to formulate the \emph{Quantitatively Nonblocking Supervisory Control Problem} of DES ({\em QNSCP}).

{\it Consider a DES plant modeled by a generator ${\bf G} = (Q,\Sigma_c \dot\cup \Sigma_{uc},\delta,q_0,Q_m)$, a specification language $E \subseteq \Sigma^*$, and let  $K := E\cap L_m({\bf G})$,
{$\mathcal{Q}_{\bf G} = $ $ \{Q_{m,i} \subseteq Q_m |i \in \mathcal{I}\}$ a partition on $Q_m$, and $N_i$ a positive integer associated with each $Q_{m,i} \in \mathcal{Q}_{\bf G}$.}
Construct a (marking) supervisory control $V: L({\bf G})\rightarrow \Gamma$ (for ($K,{\bf G}$))
satisfying the following properties:
\begin{itemize}
\item[$\bullet$]  {\bf Safety}. Marked behavior of the closed-loop system $V/{\bf G}$ satisfies the imposed specification $E$ in the sense that {$L_m(V/{\bf G}) \subseteq E\cap L_m({\bf G}) (=K)$}.

\item[$\bullet$] {\bf Quantitative nonblockingness}. Supervisory control $V$ is quantitatively nonblocking wrt. $\{({Q_{m,i}}, N_i)|i \in \mathcal{I}\}$.

\item[$\bullet$] {\bf Maximal permissiveness}. Supervisory control $V$ does not restrict more behavior than necessary to satisfy
safety and  quantitative nonblockingness, i.e. for all
other safe and quantitatively nonblocking supervisory controls $V'$ it holds that  $L_m(V'/{\bf G}) \subseteq L_m(V/{\bf G})$.
\end{itemize}
}

\begin{remark}
The QNSCP is a generalization of the the traditional nonblocking supervisory control problem \cite{RamWon87,WonRam87,Wonham16a}, in that the second requirement of quantitative nonblockingness imposes bounds on reaching subsets of marker states. This generalized problem cannot be solved in general by supervisors synthesized using the standard method; an example of ${\bf SUP}$ was given in Example~1.
\end{remark}

\begin{remark}
In \cite{QueirozEt:2005} a multitasking supervisory control problem is studied, where there are multiple tasks (modeled by colors of marker states) and each task must be completed. This requirement is  formulated as strong nonblockingness of automaton. However, bounds on the number of steps completing each task is not considered, and consequently the developed method in \cite{QueirozEt:2005} cannot be
applied to solve our problem QNSCP directly.
\end{remark}

In subsequent sections, we will develop new algorithms to design supervisors satisfying
the new requirement of quantitative nonblockingness and resolving the QNSCP.

\section{Supremal Quantitatively Completable Sublanguage and Its Computation}



Towards solving the QNSCP formulated in the preceding section,
we first present a basic result which is a counterpart to Theorem~\ref{thm:sct}.

\smallskip
\begin{thm} \label{thm:snsct}
Consider a plant generator ${\bf G} = (Q, \Sigma_c \dot\cup \Sigma_{uc},\delta,q_0,Q_m)$,
{a partition $\mathcal{Q}_{\bf G} = $ $\{Q_{m,i} \subseteq Q_m | i \in \mathcal{I}\}$ on $Q_m$, and a positive integer $N_i$ associated with each $Q_{m,i} \in \mathcal{Q}_{\bf G}$.}
Let $K\subseteq L_m({\bf G})$, $K \neq \emptyset$. There exists a quantitatively nonblocking (marking) supervisory control $V$ (for $(K, {\bf G})$) such that $L_m(V/{\bf G}) = K$ if and only if $K$ is controllable and quantitatively completable wrt. $\{({Q_{m,i}}, N_i)|i \in \mathcal{I}\}$. Moreover, if such a quantitatively nonblocking supervisory control $V$ exists, then it may be implemented by a quantitatively nonblocking generator ${\bf QSUP}$, i.e. $L_m({\bf QSUP}) = L_m(V/{\bf G})$.
\hfill $\diamond$
\end{thm}

Theorem~\ref{thm:snsct} asserts that when the $K$-synthesizing supervisory control $V$ is required to be quantitatively nonblocking, it is necessary and sufficient to require that $K$ be not only controllable but also quantitatively completable.
This result extends the standard one of supervisory control theory (i.e. Theorem~\ref{thm:sct}) \cite{RamWon87,WonRam87,Wonham16a}.

If $K$ is indeed controllable and quantitatively completable, then the supervisory control $V$ in Theorem~\ref{thm:snsct} is the solution to the QNSCP. If $K$ is either not controllable or not quantitatively completable, then to achieve the third requirement of maximal permissiveness of QNSCP, one would hope that the supremal controllable and quantitatively completable sublanguage of $K$ exists. Again the key is to investigate if for quantitative completability the supremal element also exists. We provide a positive answer below. Before we proceed, the following is a proof of Theorem~\ref{thm:snsct}.

{\it Proof of Theorem~\ref{thm:snsct}.}
We first prove the first statement. The direction of (only if) is a direct result from Theorem~\ref{thm:sct} and Definition~\ref{defn:snnbs}.
For the direction of (if), according to Theorem~\ref{thm:sct}, since $K$ is controllable, there exists a supervisory control $V$ such that
$V$ is nonblocking and $L_m(V/{\bf G}) = K$. Furthermore, according to Definition~\ref{defn:snnbs}, it is derived from $L_m(V/{\bf G}) = K$ being quantitatively completable wrt. $\{({Q_{m,i}}, N_i)|i \in \mathcal{I}\}$ that  $V$ is quantitatively nonblocking wrt. $\{({Q_{m,i}}, N_i)|i \in \mathcal{I}\}$.

For the second statement, let $V$ be a quantitatively nonblocking supervisory control that synthesizes a controllable and quantitatively completable $K$, i.e. $L_m(V/{\bf G}) = K$.
{
Since $K$ is quantitatively completable,
it follows from Proposition~\ref{prop:relation} that there exists a quantitatively nonblocking {\bf QSUP} such that $L_m({\bf QSUP}) = K = L_m(V / {\bf G})$.
} This completes the proof.
\hfill $\blacksquare$

By Theorem~\ref{thm:snsct}, the solvability of QNSCP is characterized by two language properties: controllability and quantitative completability. For controllability, it is well known that this property is closed under union, and thus there exists the supremal controllable sublanguage of a given language. We show that the same algebraic well-behavedness is enjoyed by quantitative completability in the subsection below.


\subsection{Supremal Quantitatively Completable Sublanguage}

{For the time being, we put aside controllability and focus on quantitative completability of languages. In particular, we will develop a method to compute the supremal quantitatively completable sublanguage.}  We first present the following proposition that quantitative completability is closed under set unions.

\begin{prop} \label{pro:sncomp}
Consider a generator ${\bf G} = (Q, \Sigma,\delta,q_0,Q_m)$,
{a partition $\mathcal{Q}_{\bf G} = \{Q_{m,i} \subseteq Q_m | i \in \mathcal{I}\}$ on $Q_m$, and a positive integer $N_i$ associated with each $Q_{m,i} \in \mathcal{Q}_{\bf G}$.}
Let $K_1, K_2 \subseteq L_m({\bf G})$.
If both $K_1$ and $K_2$ are quantitatively completable wrt. $\{({Q_{m,i}}, N_i)|i \in \mathcal{I}\}$,
then $K := K_1\cup K_2$ is also quantitatively completable wrt. $\{({Q_{m,i}}, N_i)|i \in \mathcal{I}\}$.
\end{prop}

{\it Proof:} 
Let $s \in \overline{K}$ and {$i \in \mathcal{I}$.}
According to Definition~\ref{defn:sncomp}, to show that $K$ is quantitatively completable, we need to show that (i) $M_{K,i}(s) \neq \emptyset$, i.e. there exists $t \in \Sigma^*$ such that {$st \in K_i = K \cap L_{m,i}({\bf G})$}, and (ii) for all {$t \in M_{K,i}(s)$}, $|t| \leq N_i$. Since $\overline{K} = \overline{K_1 \cup K_2} = \overline{K_1} \cup \overline{K_2}$, either $s \in \overline{K}_1$ or $s \in \overline{K}_2$. We consider the case $s \in \overline{K}_1$; the other case is similar.

We first show that (i) holds. Since $K_1$ is quantitatively completable, $M_{K_1,i}(s) \neq \emptyset$, i.e. there exists
string $t$ such that {$st \in K_1 \cap L_{m,i}({\bf G}) \subseteq K \cap L_{m,i}({\bf G})$}. Thus (i) is established.

For (ii), let {$t \in M_{K,i}(s)$}; then {$st \in K\cap L_{m,i}({\bf G})$} and for all $t'\in \overline{t} \setminus\{t\}$, {$st' \notin K\cap L_{m,i}({\bf G})$}.
Since $K = K_1 \cup K_2$, there exist the following two cases: (a) {$st \in K_1 \cap L_{m,i}({\bf G}) $} and for all $t'\in \overline{t} \setminus\{t\}$, {$st' \notin K\cap L_{m,i}({\bf G})$};
(b) {$st \in K_2 \cap L_{m,i}({\bf G})$} and for all $t'\in \overline{t} \setminus\{t\}$, {$st' \notin K \cap L_{m,i}({\bf G})$}.
For case (a), it follows from $K \supseteq K_1$ that {$st' \notin K_1\cap L_{m,i}({\bf G})$}, so {$t \in M_{K_1,i}(s)$}. Since $K_1$ is quantitatively completable, it holds that $|t| \leq N_i$.
The same conclusion holds for case (b) by a similar argument on $K_2$. Hence (ii) is established.

 With (i) and (ii) as shown above, we conclude that $K$ is quantitatively completable. \hfill $\blacksquare$
\smallskip

{Following an analogous proof as above, it can be shown that quantitative completability is closed under arbitrary set unions. Namely if each $K_\alpha$ of $\{K_\alpha | \alpha \in A\}$ ($A$ an index set) is quantitatively completable wrt. $\{({{Q}_{m,i}}, N_i)|i \in \mathcal{I}\}$, then $K = \bigcup_{\alpha \in A} K_\alpha$ is also quantitatively completable wrt. $\{({{Q}_{m,i}}, N_i)|i \in \mathcal{I}\}$.}

Now for given a sublanguage $K \subseteq L_m({\bf G})$, whether or not $K$ is quantitatively completable wrt. $\{({{Q}_{m,i}}, N_i)|i \in \mathcal{I}\}$, let
\begin{align*}
{\mathcal{QC}}&(K, \{({ Q_{m,i}}, N_i) | i \in \mathcal{I}\}):= \{K' \subseteq K \mid K' \notag\\
 &\text{is quantitatively completable wrt.}\{({Q_{m,i}}, N_i)|i \in \mathcal{I}\}\} \notag\\
\end{align*}
represent the set of sublanguages of $K$ that are quantitatively completable wrt. $\{({{Q}_{m,i}}, N_i)|i \in \mathcal{I}\}$.
Note from Definition \ref{defn:sncomp} that the empty language $\emptyset$ is trivially quantitatively completable, so $\emptyset \in $ ${\mathcal{QC}}(K, \{({ Q_{m,i}}, N_i) | i \in \mathcal{I}\})$ always holds. 
Moreover, it follows from Proposition~\ref{pro:sncomp} that there exists the supremal quantitatively completable sublanguage of $K$ wrt. $\{({{Q}_{m,i}}, N_i)|i \in \mathcal{I}\}$, given by
\begin{align*}
\sup\mathcal{QC}(K, &\{({Q_{m,i}}, N_i) | i \in \mathcal{I}\}) := \bigcup\{ K' \mid \\
&~~~~~~~~~~~~~~~K' \in \mathcal{QC}(K, \{({Q_{m,i}}, N_i) | i \in \mathcal{I}\}) \}.
\end{align*}

{
To compute this $\sup\mathcal{QC}(K, \{({Q_{m,i}}, N_i) | i \in \mathcal{I}\})$, we proceed as follows.
Fix $i \in \mathcal{I}$ and let
\begin{align*}
{\mathcal{QC}}&(K, ({ Q_{m,i}}, N_i)):= \{K' \subseteq K \mid K' \notag\\
 &~~~~~~ \text{is quantitatively completable wrt.}({Q_{m,i}}, N_i)\}
\end{align*}
be the set of all quantitatively completable sublanguage of $K$ wrt. $({Q_{m,i}}, N_i)$ (Definition \ref{defn:sncomp}).
By the same reasoning as above, we have that
$\sup\mathcal{QC}(K, ({Q_{m,i}}, N_i))$ exists. The idea of our algorithm design is to first compute $\sup\mathcal{QC}(K, ({Q_{m,i}}, N_i))$ for a fixed $i \in \mathcal{I}$, and then iterate over all $i \in \mathcal{I}$ until fixpoint in order to compute $\sup\mathcal{QC}(K, \{({Q_{m,i}}, N_i) | i \in \mathcal{I}\})$.
}

\subsection{Computation of $\sup\mathcal{QC}(K, ({Q_{m,i}}, N_i))$}

Consider a language $K \subseteq L_m({\bf G})$ and $({Q_{m,i}}, N_i)$ for a fixed $i \in \mathcal{I}$.
In the subsection, we present a language formula for $\sup\mathcal{QC}(K, ({Q_{m,i}}, N_i))$.

To this end, we introduce several {notations}. {For integer $N_i$, let $\Sigma^{N_i}$ be the set of strings in $\Sigma^*$ that have
lengths no more than $N_i$, i.e. $\Sigma^{N_i} := \{t \in \Sigma^*| ~|t| \leq N_i\}$.
Next, for language $K\subseteq L_m({\bf G})$, subset of marker states $Q_{m,i}$ and integer $N_i$, let
\begin{align} \label{eq:Ki}
K_i := K \cap L_{m,i}({\bf G}).
\end{align}
Then define
\begin{align} \label{eq:K'}
\widetilde{K_i} := \overline{K_i} \cap (\Sigma^{N_i-1} \cup K_i \Sigma^{N_i})
\end{align}
where $K_i\Sigma^{N_i} := \{st|s\in K_i ~\&~ t \in \Sigma^{N_i}\}$.
In simple words, $\widetilde{K_i}$ contains two subsets of $\overline{K_i}$: the first subset includes the strings that have length no more than $N_i-1$.
The second subset includes the strings each of which is a catenation of a string in $K_i$ and a string having length no more than $N_i$.

Now let
\begin{align} \label{eq:K''}
pre(\widetilde{K_i}) := \{s \in \Sigma^* | \overline{s} \subseteq \widetilde{K_i}\}.
\end{align}
Note that $pre(\widetilde{K_i})$ is prefix-closed, i.e. $pre(\widetilde{K_i}) = \overline{pre(\widetilde{K_i})}$. To see this, first the direction $(\subseteq)$ is automatic (by definition of prefix-closedness). For the reverse direction $(\supseteq)$, let $s\in \overline{pre(\widetilde{K_i})}$; then there exists $t \in \Sigma^*$ such that $st \in pre(\widetilde{K_i})$. So by (\ref{eq:K''}), we have $\overline{st} \subseteq \widetilde{K_i}$. Furthermore, it follows from ${\overline{s}} \subseteq {\overline{st}}$ that $\overline{s} \subseteq \widetilde{K_i}$, and therefore $s \in pre(\widetilde{K_i})$.


Based on (\ref{eq:K''}), we can find all the prefixes of strings in $\overline{K_i}$ that lead a string from $\overline{K_i}\setminus K_i$
to $K_i$ in no more than $N_i$ steps. As will be confirmed by the following theorem, by the computation of
$pre(\widetilde{K_i})$, we can find the supremal quantitatively completable sublanguage of $K$ with respect to $(Q_{m,i}, N_i)$, i.e. $\sup\mathcal{QC}(K, ({Q_{m,i}}, N_i))$.

\begin{thm} \label{thm:languagecompute}
Given a language $K \subseteq L_m({\bf G})$,  a subset $Q_{m,i} \subseteq Q_m$ of marker states and a positive integer $N_i$, let $\widetilde{K_i}$ and $pre(\widetilde{K_i})$ be the languages defined
in (\ref{eq:K'}) and (\ref{eq:K''}) respectively. Then,
\begin{align} \label{eq:languagecompute}
\sup\mathcal{QC}(K, ({Q_{m,i}}, N_i)) = pre(\widetilde{K_i}) \cap K.
\end{align}
\end{thm}

By the above theorem, $\sup\mathcal{QC}(K, ({Q_{m,i}}, N_i))$ can be expressed by the formula (\ref{eq:languagecompute}), and thus can be computed
by the operations on languages (union, intersection, catenation) as expressed by formulas (\ref{eq:Ki})-(\ref{eq:languagecompute}).
In particular, (\ref{eq:Ki}),
(\ref{eq:K'}), and (\ref{eq:languagecompute}) can be implemented by the product of generators representing languages $K$, $L_{m,i}({\bf G})$, $\overline{K_i}$,
$\Sigma^{N_i-1}$ and $K_i\Sigma^{N_i}$, and (\ref{eq:K''}) can be
implemented by removing the non-marker states of the automaton representing $\widetilde{K_i}$ which in turn need generators representing languages $\overline{K_i}$,
$\Sigma^{N_i-1}$, and $K_i\Sigma^{N_i}$. Thus the key is to construct two generators representing
$\Sigma^{N_i-1}$ and $K_i\Sigma^{N_i}$, respectively (generators representing $K$, $L_{m,i}({\bf G})$, and $\overline{K_i}$ are readily constructible).

First, for $\Sigma^{N_i-1}$, we construct ${\bf A}_1 = (Y_1, \Sigma, \eta_1, y_{1,0}, Y_1)$ with $Y_1 = \{y_{1,0}, y_{1,1}, ..., y_{1,N_i-1}\}$, and $\eta_1(y_{1,i},\sigma) = $ $y_{1,i+1}$ for all $\sigma \in \Sigma$ and $0\leq i\leq N_i-2$. It is easily verified that $L_m({\bf A}_1) = \Sigma^{N_i-1}$.

Second, since $K_i\Sigma^{N_i}$ is the catenation of two languages $K_i$ and $\Sigma^{N_i}$, a standard method \cite{Hopcroft14} is to first construct two generators ${\bf B}_1$ and ${\bf B}_2$ representing
$K_i$ and $\Sigma^{N_i}$ respectively and then add $\epsilon$-transitions between
the marker states of ${\bf B}_1$ and the initial state of ${\bf B}_2$. However, this combined generator is non-deterministic, and transforming it
into a deterministic generator is exponential in the state size of the combined generator in the worst case.
More precisely, it is shown in \cite{Yuzhuang:1994,Jirasek:2005} that the complexity of
computing the catenation $K_i\Sigma^{N_i}$ is $O((2m-k)2^{n-1})$, where $m$ and $k$ are respectively the numbers of the states and marker states of ${\bf B}_1$, and $n$ is the number of states of ${\bf B}_2$. Since $n=N_i+1$ according to the construction of ${\bf A}_1$ above, the complexity of computing $K_i\Sigma^{N_i}$ is exponential in $N_i$. Hence, based purely on language operations, the complexity of computing $pre(\widetilde{K_i})$ and $\sup\mathcal{QC}(K, ({Q_{m,i}}, N_i))$ is exponential in $N_i$.

In the following, we present a generator-based algorithm to compute the language  $pre(\widetilde{K_i})$ which is {\em polynomial} in $N_i$ as well as in the state numbers of the involved generators. As a result, we present a polynomial algorithm that computes $\sup\mathcal{QC}(K, $ $({Q_{m,i}}, N_i))$.

The idea of this generator-based algorithm is as follows.
Since $\widetilde{K_i}$ includes all the strings in $\overline{K_i}$ that lead a string from $\overline{K_i}\setminus K_i$
to $K_i$ in no more than $N_i$ steps, it suffices to find for each string of $\overline{K_i}$ the quantitatively completable strings, and remove
other non-quantitatively completable strings from $\overline{K_i}$.
Following this idea, a generator can be directly constructed to represent the language $pre(\widetilde{K_i})$. The detailed steps are described in Algorithm~\ref{algm:supKi} below. In the algorithm,
we employ a last-in-first-out stack $ST$ to store the states to be processed (a first-in-first-out queue can also be used instead to perform a different order of search), and for a set $Z$ a flag $F:Z \rightarrow \{true, false\}$ to
indicate whether or not an element of $Z$ has been visited: $F(z) = true$ represents that $z \in Z$ has been visited.

\begin{algorithm}[htbp]{
\caption{: Algorithm of Computing $pre(\widetilde{K_i})$}
\label{algm:supKi}

\noindent {\bf Input}:  Language $K_i$ and positive integer $N_i$.\\
\noindent {\bf Output}: Generator ${\bf K}_i'= (X_i', \Sigma, \xi_i', x'_{i,0}, X_{i,m}')$.
\\
\\
\noindent {\bf Step 1.} Construct a generator ${\bf K}_i = (X_i, \Sigma, \xi_i, x_{i,0},$ $ X_{i,m})$ to represent $K_i$, and let
\[X_i' := \{(x_i,d)|x_i\in X_i, d \in \{0,...,N_i-1\}\},\]
$\xi_i' = \emptyset$, $x_{i,0}' = (x_{i,0},0)$, and $X_{i,m}' := \{(x_i,0)|x_i\in X_{i,m}\}$. Initially set $F((x_i,d)) = false$ for each state $x_i \in X_i$ and each $d \in \{0,...,N_i-1\}$.
Then push the initial state $x_{i,0}'=(x_{i,0},0)$ into stack $ST$, and set $F((x_{i,0},0)) = true$.
\\
\\
\noindent {\bf Step 2.} If stack $ST$ is empty, trim\footnotemark the generator ${\bf K}_i' = (X_i', \Sigma, \xi_i', x'_{i,0}, X_{i,m}')$,
and output the trimmed automaton ${\bf K}_i'$ with $X_{i,m}' = X_i'$.
Otherwise, pop out the top element $(x_{i,j}, d)$ of stack $ST$. If $x_{i,j} \in X_{i,m}$, go to Step 3; otherwise, go to Step 4.
\\
\\
\noindent {\bf Step 3.} For each event $\sigma \in \Sigma$ defined at state $x_{i,j}$ (i.e. $\xi(x_{i,j},\sigma)!$), let $x_{i,k} := \xi(x_{i,j},\sigma)$ and do the following two steps 3.1 and 3.2;
then go to Step 2 with updated stack $ST$.\\
\\
\indent~~ {\bf Step 3.1} Add transition $((x_{i,j},0),\sigma,(x_{i,k},0))$ to $\xi'$, i.e.
\[\xi_i' := \xi_i' \cup \{((x_{i,j},0),\sigma, (x_{i,k},0))\}.\]
\indent~~ {\bf Step 3.2} If $F((x_{i,k},0)) = false$, push $(x_{i,k},0)$ into stack $ST$ and set $F((x_{i,k},0)) = true$.
\\
\\
\noindent {\bf Step 4.} For each event $\sigma \in \Sigma$ defined at state $x_{i,j}$ (i.e. $\xi(x_{i,j},\sigma)!$), do the following three steps 4.1--4.3;
then go to Step 2 with updated stack $ST$.
\\
\indent~~ {\bf Step 4.1} Let $x_{i,k} := \xi_i(x_{i,j},\sigma)$. If $x_{i,k} \in X_{i,m}$, set $d' = 0$; if $x_{i,k} \notin X_{i,m}$, set $d' = d + 1$.
\\
\indent~~ {\bf Step 4.2} If $d' = N_i$, go to Step 4.1 with the next event $\sigma$ defined at $x_{i,j}$.
Otherwise, add a new transition $((x_{i,j},d),\sigma, (x_{i,k},d'))$ to $\xi'$, i.e.
\[\xi_i' := \xi_i' \cup \{((x_{i,j},d),\sigma, (x_{i,k},d'))\}\]
\indent~~ {\bf Step 4.3} If $F((x_{i,k},d')) = false$, push $(x_{i,k},d')$ into stack $ST$ and set $F((x_{i,k},d')) = true$.
}
\end{algorithm}
\footnotetext{`Trimmed' means that all non-reachable and non-coreachable states (if they exist) are removed \cite{Elienberg:1974, Wonham16a}.
The generator ${\bf K}'$ need not be trim, and to get a nonblocking generator, this step of trimming is required.  }

In Step 4.2, note that the condition $d' = N_i$ means that the $N_i$-step downstream transitions that have never
reached a marker state in $Q_{m,i}$ will be removed, therefore guaranteeing that from an arbitrary state, at most $N_i$-step transitions
are needed to reach a marker state in $Q_{m,i}$.

Now we present an example to illustrate Algorithm~\ref{algm:supKi}.

\begin{example} [Continuing Example 1]
As in Fig.~\ref{fig:GSUP} consider generator ${\bf G}$, language $K = L_m({\bf SUP})$, marker state subset ${Q_{m,1}} = \{1,2\} \subseteq Q_m$ and
positive integer $N_1 = 3$. First, compute $L_m({\bf SUP}_1) := $ $L_m({\bf SUP}) \cap L_{m,1}({\bf G})$ (represented by generator ${\bf SUP}_1$ as displayed in Fig.~\ref{fig:SUP1}).
Then, inputting $K_1 = L_m({\bf SUP}_1)$ and $N_1 = 3$, Algorithm~\ref{algm:supKi} outputs a new language $pre(\widetilde{K_1})$ (represented by ${\bf NSUP}_1$ with all states marked) as displayed in Fig.~\ref{fig:NSUP1}.
By construction, every string $s$ in $L_m({\bf NSUP}_1)$ visiting the marker state (1,0) (resp. the marker state (2,0)) visits
the marker state 1 (resp. the marker state 2) in $Q_{m,1}$ of ${\bf G}$. Thus, marker state (1,0) (resp. marker state (2,0)) of ${\bf NSUP}_1$ {\it corresponds} to marker state 1 (resp. marker state 2) of ${\bf G}$.
Note also that the reason why all states in ${\bf NSUP}_1$ are marked is because ${\bf NSUP}_1$ represents the set of all the prefix strings that can be extended to reach states (1,0) and (2,0) (corresponding to marker states 1 and 2 in $Q_{m,1}$ of {\bf G}, respectively) in at most 3 steps.

According to formula (\ref{eq:languagecompute}) in Theorem~\ref{thm:languagecompute}, by the intersection of language $L_m({\bf NSUP}_1)$ and $L_m({\bf SUP})$, we obtain the supremal quantitatively completable sublanguage wrt. $(Q_{m,1}, N_1) = (\{1,2\}, 3)$.
\hfill $\diamond$

\begin{figure}[!t]
\centering
    \includegraphics[scale = 0.4]{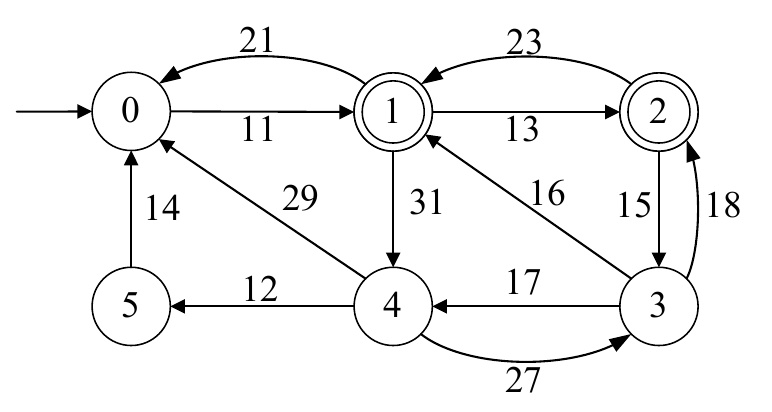}
\caption{Transition graph of ${\bf SUP}_1$ representing $K_1$} \label{fig:SUP1}
\vspace{-0.2cm}
\end{figure}

\begin{figure}[!t]
\centering
    \includegraphics[scale = 0.5]{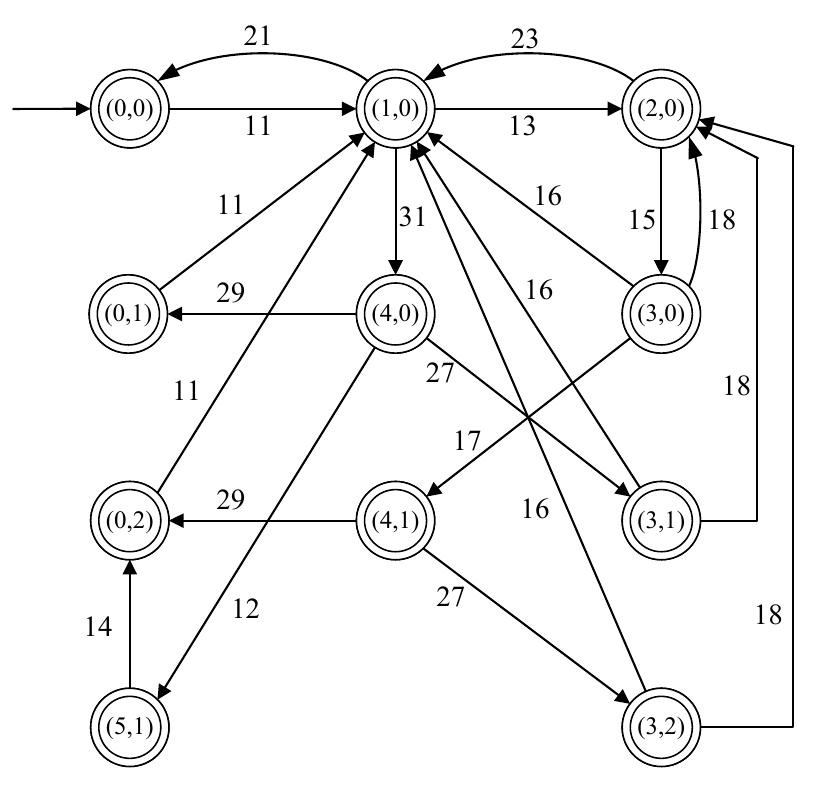}
\caption{Transition graph of ${\bf NSUP}_1$ representing $pre(\widetilde{K_1})$} \label{fig:NSUP1}
\vspace{-0.2cm}
\end{figure}

\end{example}

The correctness of Algorithm 1 is confirmed by the following proposition.

\begin{prop} \label{prop:algmcompute}
Given a language $K_i$ and a positive integer $N_i$, let ${\bf K}_i'$ be the generator returned by Algorithm~1. Then
$L_m({\bf K}_i') = pre(\widetilde{K_i})$.
\end{prop}

The above theorem confirms that Algorithm 1 computes a generator representing $pre(\widetilde{K_i})$.
The time complexity of Algorithm 1 is $O(|X_i|\cdot|\Sigma|\cdot N_i)$, where $|X_i|$ is the state number of automaton
${\bf K}_i$. This complexity is derived according to Steps 3 and 4 in Algorithm~1, because
${\bf K}_i$ has at most $|X_i| \cdot |\Sigma|$ transitions and each transition is visited at most $N_i-1$ times.
Hence, Algorithm 1 based on generators is a polynomial algorithm for computing $pre(\widetilde{K_i})$, in contrast to the language operations for $pre(\widetilde{K_i})$ of worst-case exponential complexity in $N_i$. 

Now, we present the proofs of Theorem~\ref{thm:languagecompute} and Proposition~~\ref{prop:algmcompute}.

{\it Proof of Theorem~\ref{thm:languagecompute}}: For simplicity in notation, let $K_i' := pre(\widetilde{K_i}) \cap K$ in this proof.
First, we prove that $K_i' \in \mathcal{QC}(K, ({Q_{m,i}}, N_i))$.
Since $K_i' = pre(\widetilde{K_i}) \cap K $ $ \subseteq K$ and the empty language is trivially quantitatively completable wrt. $(Q_{m,i}, N_i)$,  we only need to show that when $K_i'$ is nonempty, it is quantitatively completable wrt. $(Q_{m,i}, N_i)$.

Let $s \in \overline{K_i'}$, $t\in \Sigma^*$, and suppose $t \in M_{K_i',i}(s)$ (since $K_i' \neq \emptyset$ and $s \in \overline{K_i'}$, we have $M_{K_i',i}(s) \neq \emptyset$).
According to Definition~\ref{defn:sncomp}, to show that $K_i'$ is quantitatively completable wrt. $(Q_{m,i}, N_i)$, we will show that $|t| \leq N$.

Since $s \in \overline{K_i'} \subseteq \overline{pre(\widetilde{K_i})} \cap \overline{K}$, we have
$s \in \overline{pre(\widetilde{K_i})}$ and $s \in \overline{K}$. Since $t \in M_{K_i',i}(s)$, we have $st \in K_i' = $ $pre(\widetilde{K_i}) \cap K$, i.e.
$\overline{st} \subseteq \widetilde{K_i}$ and $st \in K$. By $\overline{st} \subseteq \widetilde{K_i}$, we have $st \in \widetilde{K_i}$, and for all prefix $t' \in \overline{t}\setminus \{t\}$, $st' \in \widetilde{K_i}$; also according to the definition of $M_{K_i',i}$, $st' \notin K_i'$. According to (\ref{eq:K'}), it derives from $st\in \widetilde{K_i}$ that $st \in \overline{K_i} \cap \Sigma^{N_i-1}$ or $st \in $ $\overline{K_i} \cap K_i\Sigma^{N_i}$.
In the former case, it holds that $|t| \leq N_i$ directly. In the latter case, we assume on the contrary that $|t| > N_i$, namely $t$ can be written
as $t = \sigma_{N_i+j}...\sigma_{N_i}...\sigma_{1}$ where $j > 1$. Then, since $\sigma_{N_i+j}...\sigma_{N_i}...\sigma_{2} \in \overline{t}$, $s\sigma_{N_i+j}...\sigma_{N_i}...\sigma_{2}\in \overline{K_i} \cap $ $K_i\Sigma^{N_i}$, which implies that there must exist an $l \geq 2$ and $l \leq N_i+1$
such that $s\sigma_{N_i+j}...\sigma_l \in K_i \subseteq K$, and thus $s\sigma_{N_i+j}...\sigma_l $ $\in K\cap pre(\widetilde{K_i}) = K_i'$. However, this contradicts the condition that for all $t' \in \overline{t}\setminus\{t\}, st' \notin K_i'$. Thus, we conclude that $|t| \leq N_i$.

It remains to show that $K_i' = pre(\widetilde{K_i}) \cap K$ is the supremal sublanguage $\sup \mathcal{QC}(K, ({Q_{m,i}}, N_i))$.
Let $M \in \mathcal{QC}(K, ({Q_{m,i}}, N_i))$ be another sublanguage of $K$ (i.e. $M\subseteq K$) that is quantitatively completable wrt. $(Q_{m,i}, N_i)$. It will be shown that  $M \subseteq K_i'$.
Namely, for any $s \in M$, we show that $s \in K_i'$.

Since $M \subseteq K$ is quantitatively completable wrt. $(Q_{m,i}, N_i)$, we have for all $s' \in \overline{M}$,
$M_{M,i}(s') \neq \emptyset$; namely, there exists $t \in \Sigma^*$ such that $s't \in M \cap L_{m,i}({\bf G}) \subseteq K \cap L_{m,i}({\bf G}) = K_i$.
So for $s \in M$, we have $\overline{s}\subseteq \overline{K \cap L_{m,i}({\bf G})} = \overline{K_i}$,
and thus we only need to prove that $\overline{s} \subseteq \Sigma^{N_i-1} \cup K_i\Sigma^{N_i}$. Let $|s| = k$ where $k \geq 0$, and write $s = \sigma_{1}...\sigma_k$. If $k \leq N_i - 1$,
it immediately follows that $\overline{s} \subseteq \Sigma^{N_i-1}$. 
So we prove in the following that
if $k \geq N_i$, then $\overline{s} \subseteq \Sigma^{N_i-1} \cup K_i\Sigma^{N_i}$.

First, we claim that there must exist a string $s_0 = \sigma_{1}...\sigma_{k_0}$ with $k_0 \leq N_i$ such that $s_0 \in M$ and $s_0 \in L_m({\bf G}_i)$;
otherwise string $s_0 \in M_{M,i}(\epsilon)$, but $|s_0| > N_i$, which implies that $M$ is not quantitatively completable wrt. $(Q_{m,i}, N_i)$
(hence a contradiction).
By $s_0 \in M \subseteq K$ and $s_0 \in L_{m,i}({\bf G})$, we have $s_0 \in K_i = K \cap L_{m,i}({\bf G})$. Then, string $s$ can be written as $s = s_0 \sigma_{k_0+1}...\sigma_k$.
On the one hand if $k \leq k_0 + N_i$, then $s \in K_i\Sigma^{N_i}$; on the other hand, since $M$ is quantitatively completable wrt. $(Q_{m,i}, N_i)$,
by the same reason as above there must exist a string $s_1 = s_0\sigma_{k_0+1}...\sigma_{k_0+k_1}$ with
$k_1 \leq N_i$ such that $s_1 \in M$ and $s_1 \in L_{m,i}({\bf G})$.

Since string $s$ is finite, $s$ can be written as $s = s_m \sigma_{k_m+1}...\sigma_k$ for some finite integer $m$, with $k \leq k_0+k_1+...+k_m+N_i$. Repeating the above process, we have $s_m \in K_i$, and thus $s \in K_i\Sigma^{N_i}$, which derives that $\overline{s} \subseteq K_i\Sigma^{N_i}$. Finally, we conclude that for all $s \in M$, we have
$s \in K \cap pre({\widetilde{K_i}})$.
The proof is now complete. \hfill $\blacksquare$

{\it Proof of Proposition~\ref{prop:algmcompute}}:
When $K_i = \emptyset$, we have $L_m({\bf K}_i') = pre(\widetilde{K_i}) = \emptyset$.
Also, if $N_i = 0$, on the one hand, $\widetilde{K_i} = \overline{K_i} \cap K_i$, and thus $pre(\widetilde{K_i})$ is the prefix-closed sublanguage of $K_i$,
which is represented by the subautomaton of ${\bf K}_i$ that contains only the marker states of ${\bf K}_i$;
on the other hand, since $N_i = 0$, Step 4.2 will not be executed, and thus all the non-marker states in ${\bf K}_i$ will be removed. So in this case we have $L_m({\bf K}_i') = L({\bf K}_i') = pre(\widetilde{K_i})$.

In the following we consider the case that $K_i \neq \emptyset$ and $N_i > 0$.
According to Step 2 in Algorithm 1, it follows from $X_{i,m}' = X_i'$ that $L({\bf K}_i') = L_m({\bf K}_i')$.
First, we prove that $L_m({\bf K}_i') = L({\bf K}_i') \subseteq pre(\widetilde{K_i})$
by induction on the length of a string $s \in L({\bf K}_i')$.

{\bf Base case:} Let $s = \epsilon \in L({\bf K}_i')$. We have $\epsilon \in \overline{K_i}$ ($K_i$ is nonempty) and $\epsilon \in \Sigma^{N_i - 1}$; thus $\epsilon \in pre(\widetilde{K_i})$.

{\bf Inductive case:} Let $s \in L({\bf K}_i')$, $s \in pre(\widetilde{K_i})$, $\sigma \in \Sigma$, and suppose that $s\sigma \in L({\bf K}_i')$; we will show that $s\sigma \in pre(\widetilde{K_i})$ as well.
Since $s \in L({\bf K}_i')$, there exists a state $x_{i,k} \in X_i'$ such that $\xi_i'(x_{i,0}',s) = x_{i,k}'$.
Also, since $s \in pre(\widetilde{K_i})$, we have $s \in \overline{K_i}$, and thus $\xi_i(x_{i,0}, s)!$; furthermore there exist
$x_{i,k} \in X_{i}$ and $d$ such that $ 0 \leq d \leq N_i - 1$, $\xi_i(x_{i,0},s) = x_{i,k}$, and $x_{i,k}' = (x_{i,k}, d)$.
By $s\sigma \in L({\bf K}_i')$, we derive that $\xi_i'(x_{i,k}',\sigma)$ is defined by Steps 3 or 4. In the former case,
we have $x_{i,k} \in X_m$ and thus $s\in K_i$, which implies that $s\sigma \in K_i\Sigma^{N_i}$.

In the latter case, according to Step 4.1, if $\xi_i(x_{i,k},\sigma) \in X_{i,m}$, we have $s\sigma \in K_i$ directly.
If $\xi_i(x_{i,k},\sigma) \notin X_{i,m}$, by Step 4.2, we have $d' = d+1$ and $d' < N_i$; otherwise, $\xi_i'(x_{i,k}',\sigma)$ will not be defined in $\xi_i'$.
Thus $d \leq N_i - 2$.
It follows from $s \in pre(\widetilde{K_i})$ that (i) $\overline{s} \subseteq \overline{K_i} \cap \Sigma^{N_i-1}$ or (ii)$s \notin \overline{K_i} \cap\Sigma^{N_i-1}$ but $\overline{s} \subseteq \overline{K_i}\cap(K_i\Sigma^{N_i})$.
In case (i), $s \in \Sigma^{d} \subseteq \Sigma^{N_i - 2}$. So $s\sigma \in \overline{K_i}\cap\Sigma^{N_i - 1}$.
In case (ii), since $s \notin \Sigma^{N_i-1}$ and $s \notin K_i$, according to Step 4, we have $s \in K\Sigma^d$.
Due to $d \leq N_i - 2$, we have $s\sigma \in K_i\Sigma^{d+1} \subseteq K_i\Sigma^{N_i - 1}$.

Combined with $s\sigma \in \overline{K_i}$, we have $s\sigma \in \overline{K_i} \cap (\Sigma^{N_i - 1} \cup K_i\Sigma^{N_i}) = \widetilde{K_i}$.
Furthermore, since $\overline{s} \subseteq \widetilde{K_i}$, we have $\overline{s\sigma} \subseteq \widetilde{K_i}$,
and thus we conclude that $s\sigma \in pre(\widetilde{K_i})$.

Second, we prove the converse direction that $L_m({\bf K}_i') = L({\bf K}_i') \supseteq pre(\widetilde{K_i})$
again by induction on the length of a string $s \in pre(\widetilde{K_i})$.

{\bf Base case:} Let $s = \epsilon \in pre(\widetilde{K_i})$. We have $\epsilon \in L({\bf K}_i') = L_m({\bf K}_i')$ because $K_i$ is nonempty and according to Step 2,
$L_m({\bf K}_i')$ is prefix-closed (all states in $X_i'$ are marker states).

{\bf Inductive case:} Let $s \in pre(\widetilde{K_i})$, $s \in L({\bf K}_i')$, $\sigma \in \Sigma$, and suppose that $s\sigma \in pre(\widetilde{K_i})$; we will show that $s\sigma \in L({\bf K}_i')$ as well. It follows from $s \in pre(\widetilde{K_i})$ that $s \in \overline{K}$ and $s\in \Sigma^{N_i-1} \cup K_i\Sigma^{N_i}$. Thus
there exists a state $x_{i,k} \in X_{i}$ such that $\xi_i(x_{i,0},s) = x_{i,k}$. Also, by $s \in L({\bf K}_i')$,
there exists a state $x_{i,k}' = (x_{i,k},d)$ with $0 \leq d\leq N_i-1$ such that $\xi_i'(x_{i,0}',s) = x_{i,k}'$.
Since $s\sigma \in pre(\widetilde{K_i})$, we have $\overline{s\sigma} \subseteq \widetilde{K_i} =\overline{K_i} \cap $ $(\Sigma^{N_i-1} \cup K_i\Sigma^{N_i})$, and thus $s\sigma \in \overline{K_i}$ and $s\sigma \in \Sigma^{N_i-1} \cup K_i\Sigma^{N_i}$.
It follows from $s\sigma \in \overline{K_i}$ that $\xi_i(x_{i,k},\sigma)!$.
According to $s\sigma \in \Sigma^{N_i-1} \cup K_i\Sigma^{N_i}$, we consider the following two cases: (i) $s\sigma \in \Sigma^{N_i-1}$, and (ii) $s\sigma \notin \Sigma^{N_i-1}$, but $s\sigma \in K_i\Sigma^{N_i}$.

In case (i), since $s\sigma \in \Sigma^{N_i-1}$, we have $|s\sigma| \leq N_i - 1$, and thus $d+1 \leq N_i-1$. According to
Step 2, if $x_{i,k} \in X_{i,m}$, then $((x_{i,k}, 0),\sigma, (\xi_i(x_{i,k},\sigma),0))$ will be added to $\xi_i'$ by Step 3.1.
If $x_{i,k} \notin X_{i,m}$, since $d +1 < N_i$, $((x_{i,k}, d),\sigma, $ $(\xi_i(x_{i,k},\sigma),0))$, or  $((x_{i,k}, d),\sigma, (\xi_i(x_{i,k},\sigma),d+1))$
will be added to $\xi_i'$ by Step 4.2.

In case (ii), by $s\in \Sigma^{N_i-1} \cup K_i\Sigma^{N_i}$, there also exist two cases (a) $s \in \Sigma^{N_i-1}$, and (b) $s \notin \Sigma^{N_i-1}$, but $s \in K_i\Sigma^{N_i}$.
In case (a), since $s\sigma \notin \Sigma^{N_i-1}$, we have $s\sigma \in \Sigma^{N_i}$, and thus $s\sigma \in K_i\Sigma^{N_i} \cap \Sigma^{N_i} \subseteq K_i$.
Namely, $\xi_i(x_{i,k},\sigma) \in X_{i,m}$. Then by Step 4.2, $((x_{i,k}, d),\sigma,$ $ (\xi_i(x_{i,k},\sigma),0))$ will be added to $\xi_i'$.
In case (b), according to Step 4, by $s \in L_m({\bf K}_i')$ and $x_{i,k}' = (x_{i,k}, d)$, if $d = 0$, then $s \in K_i$ or $s \in K_i\Sigma^1$:
if $s \in K_i$, according to Step 3, the transition $((x_{i,k}, 0),\sigma, (\xi_i(x_{i,k},\sigma),0))$ will be added to $\xi_i'$;
if $s \in K_i\Sigma^1$, by $s\sigma \in \Sigma^{N_i}$, we have $N_i \geq 2$, and thus $d+1 = 1 < N_i$ and according to
Step 4.2, the transition $((x_{i,k}, 0),\sigma, (\xi_i(x_{i,k},\sigma),1))$ will be added to $\xi_i'$.
If $d > 0$, then there must exist
states $x_{i,k-1}' = (x_{i,k-1}, d -1)$, ..., $x_{i,k-d}' = (x_{i,k-d}, 0)$ and $x_{i,k-d-1}' = (x_{i,k-d-1}, 0)$ such that
transitions $(x_{i,k-1},\sigma_d, x_{i,k})$, ..., $(x_{i,k-d},\sigma_{1}, x_{i,k-d+1})$
and $(x_{i,k-d-1},\sigma_{0}, x_{i,k-d})$ exist in $\xi_i$ and $x_{i,k-d-1} \in X_{i,m}$.
Assume that $\xi_i(x_{i,0},s_0) = x_{i,k-d-1}$; then $s_0 \in K_i$ and $s = s_0\sigma_0\sigma_1...\sigma_d$.
Due to $\overline{s\sigma} \subseteq K_i\Sigma^{N_i}$, we have $d+1+1 \leq N_i$, i.e. $d + 1 < N_i$.
According to Step 4.2, the transition $((x_{i,k}, d),\sigma, (\xi_i(x_{i,k},\sigma),d+1))$ will be added to $\xi_i'$.
In all the cases above, we have shown that the transition $\xi_i'((x_{i,k},d),\sigma)$ will be added to $\xi_i'$.
Hence, we conclude that $s\sigma \in L({\bf K}_i')$. Finally, by  $L({\bf K}_i') = L_m({\bf K}_i')$, it holds
that $s \in L_m({\bf K}_i')$.
\hfill $\blacksquare$

After proving Theorem~\ref{thm:languagecompute} and Proposition~~\ref{prop:algmcompute},
we summarize in Algorithm~\ref{algm:supqck} below the steps for computing $\sup\mathcal{QC}(K, (Q_{m,i}, N_i))$.

\begin{algorithm}[htbp]{
\caption{: Algorithm of Computing $\sup\mathcal{QC}(K, (Q_{m,i}, N_i))$}
\label{algm:supqck}

\noindent {\bf Input}: Generator ${\bf G} = (Q, \Sigma, \delta, q_0, Q_m)$, language $K \subseteq L_m({\bf G})$, subset $Q_{m,i} \subseteq Q_m$ of marker states, and positive integer $N_i$.\\
\noindent {\bf Output}: Language $K_i'$.
\\
\\
\noindent {\bf Step 1.} Compute $K_i = K \cap L_{m,i}({\bf G})$ as in (\ref{eq:Ki}). If $K_i = \emptyset$, output $K_i' = \emptyset$;
otherwise go to Step 2.
\\
\\
\noindent {\bf Step 2.} Apply Algorithm~\ref{algm:supKi} with input $K_i$ and $N_i$ to compute $pre(\widetilde{K_i})$ as defined in (\ref{eq:K'}) and (\ref{eq:PKL}).
\\
\\
\noindent {\bf Step 3.} Output $K_i' := K \cap pre(\widetilde{K_i})$.
}
\end{algorithm}

Note that in Step 1, if $K_i$ is empty, then all strings in $K_i$ cannot visit marker states in $Q_{m,i}$,
and thus in this case, Step 2 is unnecessary; otherwise we need Algorithm~\ref{algm:supKi} to compute
$pre(\widetilde{K_i})$.

Let $|Q|$ and $|X|$ be the state sizes of the generator ${\bf G}$ and the generator representing $K$, respectively. In Step~1, the state size $|X_i|$ of the generator representing $K_i$ is at most $|X|\cdot |Q|$. The complexity of this step is  $O(|X|\cdot |Q|)$. In Step~2, the complexity of computing $pre(K_i)$ is $O(|X|\cdot |Q|\cdot |\Sigma|\cdot N_i)$ (see the complexity analysis of Algorithm~\ref{algm:supKi} below Proposition~\ref{prop:algmcompute}). 
{Finally, in Step~3, the intersection of $K$ and ${pre(K_i)}$ can be done by unmarking those states of the generator representing ${pre(K_i)}$ that are {\em not} visited by strings in $K$. Hence the complexity of Step~3 is $O(|X|\cdot |Q|\cdot |\Sigma|\cdot N_i)$.} Thus the overall complexity of Algorithm~\ref{algm:supqck} is $O(|X|\cdot |Q|\cdot |\Sigma|\cdot N_i)$.
 Therefore Algorithm~2 is a polynomial algorithm for computing $\sup\mathcal{QC}(K, (Q_{m,i}, N_i))$.

\begin{example} [Continuing Example 1]
For the language $pre(\widetilde{K_1})$ represented by the generator ${\bf NSUP}_1$ (in Example~3 and in Fig.~\ref{fig:NSUP1}), by Step 3 of Algorithm~2, the strings both in $K$ and $pre(\widetilde{K_1})$ are
preserved in the final language $K_1'$ as represented by ${\bf SUP}_1'$ displayed in Fig.~\ref{fig:SUP1'}.
{
Here the marker states 1 and 2 of ${\bf SUP}_1'$ correspond respectively to the marker states 1 and 2 in $Q_{m,1}$ of ${\bf G}$ (i.e. every string visiting marker state 1 (resp. marker state 2) of ${\bf SUP}_1'$ also visits marker state 1 (resp. marker state 2) of ${\bf G}$).
It is readily verified that $L_m({\bf SUP}_1') \subseteq L_m({\bf SUP})$, and every state of
${\bf SUP}_1'$ can visit its marker states 1 and 2 in at most three steps. This verifies that ${\bf SUP}_1'$ enforces quantitative nonblockingness wrt. $(Q_{m,1}, N_1) = (\{1,2\}, 3)$. }
\hfill $\diamond$

\begin{figure}[!t]
\centering
    \includegraphics[scale = 0.4]{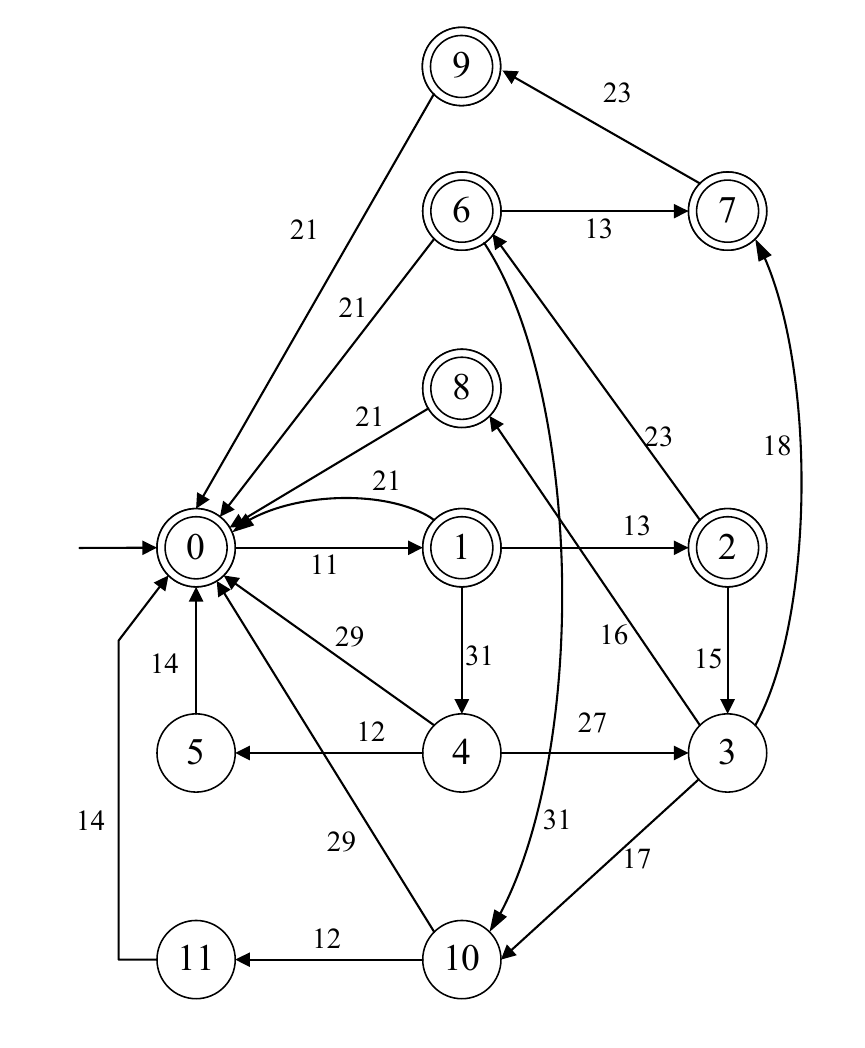}
\caption{Transition graph of ${\bf SUP}_1'$ representing $K_1'$} \label{fig:SUP1'}
\vspace{-0.2cm}
\end{figure}

\end{example}

\subsection{Computation of $\sup\mathcal{QC}(K, \{(Q_{m,i}, N_i| i \in \mathcal{I}\})$}

Now that we know how to compute $\sup\mathcal{QC}(K, (Q_{m,i}, N_i))$ for a fixed $i \in \mathcal{I}$, we proceed
to design an algorithm to compute the supremal quantitatively completable sublanguage $\sup\mathcal{QC}(K, $ $\{(Q_{m,i}, N_i| i \in \mathcal{I}\})$ by iterating over all $i \in \mathcal{I}$ until fixpoint.

Consider a generator ${\bf G} = (Q, \Sigma, \delta, q_0, Q_m)$, a sublanguage $K  \subseteq L_m({\bf G})$, a partition $\mathcal{Q}_{\bf G} =$ $ \{Q_{m,i} | i \in \mathcal{I}\}$ on $Q_m$, and a positive integer $N_i$ associated with each $Q_{m,i} \in \mathcal{Q}_{\bf G}$.
Letting $\mathcal{I} = $ $\{1,\ldots,M\}$ ($M \geq 1$), we present our algorithm of
computing $\sup\mathcal{QC}(K, \{(Q_{m,i}, N_i| i \in \mathcal{I}\})$ as follows.


\begin{algorithm}[h]{
\caption{: Algorithm of Computing $\sup\mathcal{QC}(K, $ $\{(Q_{m,i}, N_i| i \in \mathcal{I}\})$\quad ($\mathcal{I} = \{1,\ldots,M\}$)}
\label{algm:supsnc}

\noindent {\bf Input}: Generator ${\bf G} = (Q, \Sigma, \delta, q_0, Q_m)$, language $K \subseteq L_m({\bf G})$, a partition $\mathcal{Q}_{\bf G} = \{Q_{m,i} | i \in \mathcal{I}\}$
on marker state set $Q_m$, and a set of positive integers $\{N_i| i \in \mathcal{I}\}$.

\noindent {\bf Output}: Language $K'$.
\\

\noindent {\bf Step 1.} Let $j = 1$ and $K^j = K$  (i.e. $K^1 = K$).

\noindent {\bf Step 2.} Let $i = 1$. Let $K_i^j = K^j$.

\noindent {\bf Step 2.1} Apply Algorithm~\ref{algm:supqck} with inputs ${\bf G}$, $K_i^j$, $Q_{m,i}$ and $N_i$, and obtain ${NK_i^j} = \sup\mathcal{QC}(K_i^j, (Q_{m,i}, N_i))$.

\noindent {\bf Step 2.2} If $i < M$, let $K_{i+1}^j = {NK_i^j}$, advance $i$ to $i+1$ and go to Step 2.1; otherwise ($i = M$), go to Step 3.

\noindent {\bf Step 3.} Let $K^{j+1} = {NK_{M}^j}$. If $K^{j+1} = K_i^j$, output $K' = K^{j+1}$. Otherwise, advance $j$ to $j+1$ and go to {\bf Step~2}.
}
\end{algorithm}

We present an example to illustrate Algorithm~3.

\begin{example} [Continuing Example 1]
Applying Algorithm \ref{algm:supsnc} with inputs ${\bf G}$, $K = L_m(\bf SUP)$ ({\bf G} and {\bf SUP} displayed in Fig.~\ref{fig:GSUP}), $\mathcal{Q}_{\bf G} = \{Q_{m,1} = \{1,2\},$ $ Q_{m,2} = \{0\}\}$ and $\{N_1 = 3, $ $N_2 = 5\}$,
we compute the supremal quantitatively completable sublanguage of $L_m(\bf SUP)$ with respect to $\{(Q_{m,i}, N_i) | i \in \{1,2\}\}$. Here $M=2$.

At Step 1, set $K^1 = K$.
At Step 2, $K_{1}^1 = K^1 = K$.
At Step 2.1, since the inputs ${\bf G}$, $K_{1}^1$, $Q_{m,1} = \{1,2\}$ and $N_1 = 3$ of Algorithm~\ref{algm:supqck} are identical to that in Example~4, the
output language $NK_1^1$ is the same as the language represented by ${\bf SUP}_1'$ (as displayed in Fig.~\ref{fig:SUP1'}),
i.e. $NK_1^1 = L_m({{\bf SUP}_1'})$.

At Step 2.2, since $i = 1 < 2$, let $K_{2}^1 = NK_1^1$ and go to Step 2.1 with $i = 2$.
Now applying Algorithm~\ref{algm:supqck} with inputs ${\bf G}$, $K_{2}^1$, $Q_{m,2} = \{0\}$ and $N_2 = 5$,
we obtain the language $NK_2^1$ represented by ${\bf SUP}_2'$ as displayed in Fig.~\ref{fig:SUP2'}.
{From Fig.~\ref{fig:SUP2'}, it is inspected that every reachable state of ${\bf SUP}_2'$ can reach one of three maker states 0, 14 and 15 which correspond to marker state 0 in $Q_{m,2}$ of $\bf G$ (reaching marker state 0 of $\bf G$ means that the vehicle moves to zone 0) within five steps.}
Then since $i = 2 = M$, go to Step 3
and let $K^2 = NK_2^1$. Obviously $K^2 \neq K^1$, so we repeat Step 2 (including Steps 2.1 and 2.2) with $K^2$, and then
obtain $K^3$.

It is verified that $K^3 = K^2$; thus the algorithm terminates and output language $K^3$ which is represented by ${\bf SUP}_2'$ in Fig.~\ref{fig:SUP2'}.
It can be confirmed that $K^3$ is quantitatively completable wrt. $\{(Q_{m,1} = \{1,2\}, $ $N_1 = 3), (Q_{m,2} = \{0\}, N_2 = 5)\}$; {namely, every reachable state of ${\bf SUP}_2'$ can be led to one of four marker states 1, 8, 6, 9 (corresponding
to marker state 1 in $Q_{m,1}$ of {\bf G}), two marker states 2, 7 (corresponding to marker state 2 in $Q_{m,1}$ of {\bf G}) within three steps, and three marker states 0, 14, 15 (corresponding to marker state 0 in $Q_{m,2}$ of {\bf G}) within five steps.} \hfill $\diamond$

\begin{figure}[!t]
\centering
    \includegraphics[scale = 0.4]{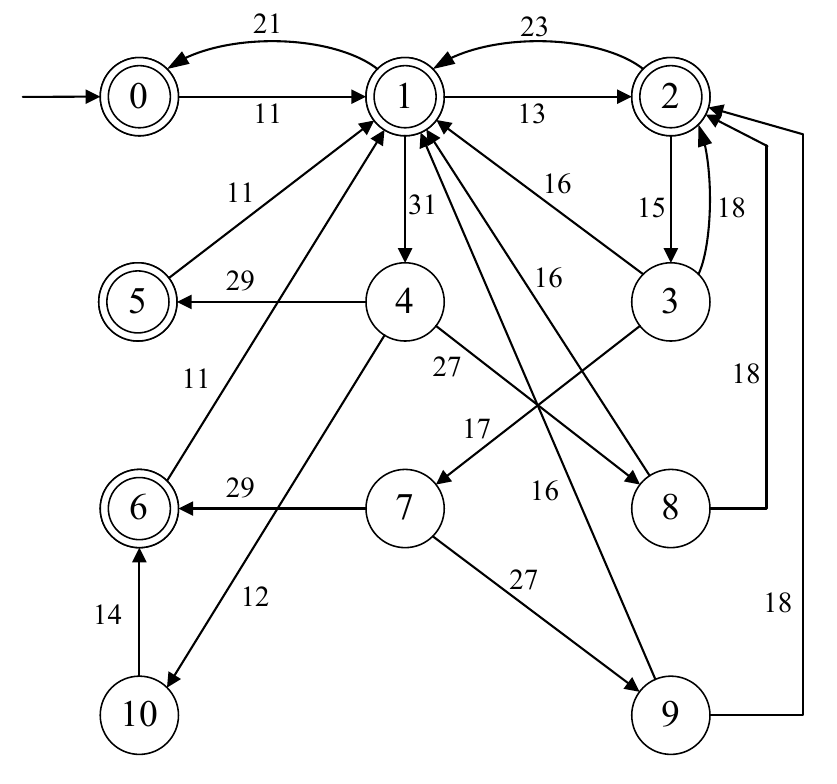}
\caption{Transition graph of ${\bf SUP}_2'$ representing $NK_2^1$ ($= K^2 = K^3$)} \label{fig:SUP2'}
\vspace{-0.2cm}
\end{figure}

\end{example}

The correctness of Algorithm~\ref{algm:supsnc} is confirmed by the following theorem.

\begin{thm} \label{thm:algmsupcomp}
Consider as inputs to Algorithm~3 a generator ${\bf G} = (Q, \Sigma, \delta, q_0, Q_m)$, a language $K \subseteq L_m({\bf G})$, a partition $\mathcal{Q}_{\bf G} = \{Q_{m,i} \subseteq Q_m | i \in \mathcal{I}\}$ on $Q_m$, and a set of positive integer $N_i$ each associated with $Q_{m,i} \in \mathcal{Q}_{\bf G}$. Then Algorithm~3 terminates in a finite number of steps and outputs a language $K'$ such that $K' = \sup\mathcal{QC}(K, \{(Q_{m,i},N_i)| i \in \mathcal{I}\})$.
\end{thm}

{\it Proof:} 
First, we prove that Algorithm~\ref{algm:supsnc} terminates in a finite number of steps. To this end, let $|X|$, $|X^j|$, $|X_i^{j}|$, $|NX_i^j|$ be the state numbers of the generators representing $K$, $K^j$, $K_i^j$, $NK_i^j$ respectively.
Initially, $|X_1^1| = |X^1| = |X|$. Then by Step 2.1, the state number $|NX_{1}^1|$ is at most $|Q|\cdot |X_{1,1}|\cdot N_1 = |Q|\cdot |X|\cdot N_1$. 
If $M (:=|\mathcal{I}|) > 1$, we continue to compute ${NK}_{2}^1$ by Step 2.1; since the the generator representing $NK_1^1$ returned by Algorithm~\ref{algm:supqck}
is constructed from the transition structure of ${\bf G}$, the state size  $|{NX}_{2}^1|$ is at most $|NX_{1}^1| \cdot N_2 = |Q|\cdot |X|\cdot N_1 \cdot N_2$.
Iterating over all $i \in \mathcal{I}$, the state size ${NX}_{M}^1 = |X^2|$ is at most $|Q|\cdot |X|\cdot N_1 \cdot N_2 \cdot ... \cdot N_{M} = |Q|\cdot |X| \cdot \prod_{i=1}^{M}{N_i}$.

When $j > 1$, since the generator representing $K^2$ is constructed from the product transition structure of generator $\bf G$, generator representing $K$, and every $N_i$ ($i \in \mathcal{I}$), the subsequent iterations when $j>1$ will not increase the state number and the transition number of the generator representing $K^{j}$. Thus the maximal state sizes of the generators in Algorithm~\ref{algm:supsnc} is:
$|Q|\cdot |X| \cdot \prod_{i=1}^{M}{N_i}$;
and the maximal transition number of the generators is $|Q|\cdot |X| \cdot |\Sigma| \cdot \prod_{i=1}^{M}{N_i}$.
In the worst case when one iteration removes only one transition, Algorithm~\ref{algm:supsnc} will terminate in $|Q|\cdot |X| \cdot |\Sigma| \cdot \prod_{i=1}^{M}{N_i}$
number of iterations, i.e. finite termination.

It is left to prove that the output $K'$ of Algorithm~3 satisfies $K' = \sup\mathcal{QC}(K, \{(Q_{m,i},N_i)| i \in \mathcal{I}\})$. Note in Algorithm~3 that Step~2 performs the computation of supremal quantitatively completable sublanguage wrt. $(Q_{m,i}, N_i)$ iteratively for each $i \in \mathcal{I}$. The iterations of Steps~2 and 3 generate the following sequence of languages:
\begin{align*}
K &= K^1 = K_1^1 \supseteq NK_1^1 \supseteq ... \supseteq NK_{M-1}^1  \supseteq NK_{M}^1\\
  &= K^2 \supseteq ...
\end{align*}
Since Algorithm~3 terminates in a finite number of steps, so does the above sequence of languages. When the sequence converges, i.e. in Step~3 $K^{j+1} = K^j$ holds for some $j$,  $K^{j+1}$ is the supremal quantitatively completable sublanguage of $K$ wrt. $\{(Q_{m,i}, N_i)| i \in \mathcal{I}\}$. This proves that $K' = K^{j+1} = \sup\mathcal{QC}(K, $ $\{(Q_{m,i},N_i)| i \in \mathcal{I}\})$.
\hfill $\blacksquare$


By the above proof, the complexity of one complete iteration over all $i \in \mathcal{I}$ in Step 2 is $O(|Q|\cdot |X| \cdot $ $\prod_{i=1}^{|\mathcal{I}|}{N_i})$.
Since there can be at most $|Q|\cdot |X| \cdot |\Sigma| \cdot $ $\prod_{i=1}^{|\mathcal{I}|}{N_i}$
iterations, the overall time complexity of Algorithm~\ref{algm:supsnc} is $O\big((|Q|\cdot |X| \cdot |\Sigma| \cdot \prod_{i=1}^{|\mathcal{I}|}{N_i})^2\big)$.

}

\section{Maximally Permissive Quantitatively Nonblocking Supervisory Control}

In this section, we present our solution to the QNSCP.
Consider a DES plant modeled by a generator ${\bf G} = (Q,\Sigma_c \dot\cup \Sigma_{uc},\delta,q_0,Q_m)$, and a specification language $E \subseteq \Sigma^*$. Let $K := E \cap L_m({\bf G})$, {$\mathcal{Q}_{\bf G} = \{Q_{m,i} | i \in \mathcal{I}\}$ be a partition on $Q_m$, and a positive integer $N_i$ associated with each $Q_{m,i} \in \mathcal{Q}_{\bf G}$.}

Whether or not $K$ is controllable and quantitatively completable, let $\mathcal{CQC}(K, {\{(Q_{m,i}, N_i)| i \in \mathcal{I}\}})$ be the set of sublanguages of $K$ that are both controllable and quantitatively completable wrt. {$\{(Q_{m,i}, N_i)| i \in \mathcal{I}\}$}, i.e.
\begin{align*}
\mathcal{CQC}&(K, {\{(Q_{m,i}, N_i)| i \in \mathcal{I}\}}) := \notag\\
&\{K' \subseteq K \mid K' ~\text{is controllable and }\notag\\
&\text{quantitatively completable wrt. ${\{(Q_{m,i}, N_i)| i \in \mathcal{I}\}}$}\}.
\end{align*}
Since the empty language $\emptyset$ is trivially controllable and quantitatively completable, the set $\mathcal{CQC}(K, $ ${\{(Q_{m,i}, N_i)| i \in \mathcal{I}\}})$ is nonempty. Moreover, since both controllability and quantitative completability are closed under arbitrary set unions, $\mathcal{CQC}(K, {\{(Q_{m,i}, N_i)| i \in \mathcal{I}\}}$ contains a unique supremal element given by
\begin{align*}
\sup\mathcal{CQC}(K, &{\{(Q_{m,i}, N_i)| i \in \mathcal{I}\}}):=\bigcup\{K' \subseteq K \mid \\
&~~~~~~~~~K' \in \mathcal{CQC}(K, {\{(Q_{m,i}, N_i)| i \in \mathcal{I}\}}) \}.
\end{align*}

Our main result in this section is the following.

\begin{thm}
Suppose that $\sup\mathcal{CQC}(K, {\{(Q_{m,i}, N_i)} {|i \in \mathcal{I}\}})\neq \emptyset$.
Then the supervisory control $V_{\sup}$ such that $L_m(V_{\sup}/{\bf G})= \sup\mathcal{CQC}(K, {\{(Q_{m,i}, N_i)| i \in \mathcal{I}\}}) $ $\subseteq K$ is the solution to the  QNSCP.
\end{thm}

{\it Proof}:
Since $\sup\mathcal{CQC}(K, {\{(Q_{m,i}, N_i)| i \in \mathcal{I}\}})$ is controllable and quantitatively completable wrt. { $\{(Q_{m,i}, N_i)| i \in \mathcal{I}\}$}, according to Theorem~\ref{thm:snsct} there exists a quantitatively nonblocking supervisory control $V_{\sup}$ such that $L_m(V_{\sup}/{\bf G})=\sup\mathcal{CQC}(K,$ $ {\{(Q_{m,i}, N_i)| i \in \mathcal{I}\}}) \subseteq K$. Hence the first (safety) and the second (quantitative nonblockingness) requirements of the QNSCP are satisfied.
Further, since $\sup\mathcal{CQC}(K, $ $ {\color{red}\{(Q_{m,i}, N_i)| i \in \mathcal{I}\}})$ is the supremal element in $\mathcal{CQC}(K, {\{(Q_{m,i}, N_i)| i \in \mathcal{I}\}})$, the third (maximal permissiveness) requirement of the QNSCP is also satisfied. Therefore, $V_{\sup}$ that synthesizes $\sup\mathcal{CQC}(K, {\{(Q_{m,i}, N_i)| i \in \mathcal{I}\}})$ is the solution to the QNSCP. \hfill $\blacksquare$

{
We proceed to design an algorithm to compute this solution $\sup\mathcal{CQC}(K, \{(Q_{m,i}, N_i)| i \in \mathcal{I}\})$.
Since there exists a well-known algorithm to compute the supremal controllable sublanguage \cite{WonRam87,Wonham16a} which
will be referred in this paper as Algorithm SC, and in the preceding section we designed Algorithm~3 to compute the supremal quantitatively completable sublanguage, a natural idea is to iterate these two algorithms until the fixed point is reached. This idea works; however, since Algorithm~\ref{algm:supsnc} is itself an iterative algorithm,
there would be two nested iterations in this approach, which would cause the overall complexity unnecessarily high.
We adopt an alternative approach, in which at the end of each iteration of Algorithm~\ref{algm:supsnc} (i.e. Step~2, 2.1, 2.2) for computing the
supremal quantitatively completable sublanguage wrt. $(Q_{m,i}, N_i)$ once for each $i \in \mathcal{I}$, we add a step to compute the supremal controllable sublanguage.
The details
are described in Algorithm~4 below.

\begin{algorithm}[h] {
\caption{: Algorithm of Computing $\sup\mathcal{CQC}(K, \{(Q_{m,i}, N_i)| i \in \mathcal{I}\})$\quad ($\mathcal{I} = \{1,\ldots,M\}$)}
\label{algm:supcsnc}

\noindent {\bf Input}: Generator ${\bf G} = (Q, \Sigma, \delta, q_0, Q_m)$, language $K \subseteq L_m({\bf G})$, partition $\mathcal{Q}_{\bf G} = \{Q_{m,i} | i \in \mathcal{I}\}$
on marker state set $Q_m$, and set of positive integers $\{N_i| i \in \mathcal{I}\}$. \\
\noindent {\bf Output}: Language $CK$.
\\

\noindent {\bf Step 1.} Let $j = 1$ and $K^j = K$ (i.e. $K^1 = K$).

\noindent {\bf Step 2.} Let $i = 1$. Let $K_i^j = K^j$.

\noindent {\bf Step 2.1} Apply Algorithm~\ref{algm:supqck} with inputs ${\bf G}$, $K_i^j$, $Q_{m,i}$ and $N_i$, and obtain ${NK_i^j} = \sup\mathcal{QC}(K_i^j, (Q_{m,i}, N_i))$.

\noindent {\bf Step 2.2} If $i < M$, let $K_{i+1}^j = {NK_i^j}$, advance $i$ to $i+1$ and go to Step 2.1; otherwise ($i = M$), go to Step 3.

\noindent {\bf Step 3.} Apply Algorithm SC with inputs ${\bf G}$ and ${NK_{M}^j}$ to compute $K^{j+1}$ such that $K^{j+1} = \sup\mathcal{C}({NK_{M}^j})$.

\noindent {\bf Step 4.} If $K^{j+1} = K^j$, output $CK = K^{j+1}$. Otherwise, advance $j$ to $j+1$ and go to {\bf Step~2}.
}
\end{algorithm}
}

The correctness of Algorithm~4 is confirmed by the following theorem.

\begin{thm}\label{thm:supstrongcomctr}
Given a plant generator ${\bf G}$, a specification language $E$, let $K := E \cap L_m({\bf G})$, {a partition $\mathcal{Q}_{\bf G} = \{Q_{m,i} \subseteq Q_m | i \in \mathcal{I}\}$ on $Q_m$, and a set of positive integer $N_i$ each associated with a $Q_{m,i} \in \mathcal{Q}_{\bf G}$}. Then Algorithm~\ref{algm:supcsnc} terminates in a finite number of steps and outputs a language $CK$ such that $CK = \sup\mathcal{CQC}(K, {\{(Q_{m,i}, N_i)| i \in \mathcal{I}\}})$.
\end{thm}

{
{\it Proof:} 
Since Algorithm~3 terminates in a finite number of steps (Theorem~\ref{thm:supstrongcomctr}), and Algorithm SC for computing the supremal controllable sublanugaes does not increase the state/transition number of the generator representing $NK_M^j$, Algorithm~\ref{algm:supcsnc} also terminates in a finite number of steps.

It is left to show that the output $CK$ of Algorithm~\ref{algm:supcsnc} satisfies $CK = \sup\mathcal{CQC}(K, {\{(Q_{m,i}, N_i)| i \in \mathcal{I}\}})$. Note that Step~2 in Algorithm~\ref{algm:supcsnc} (same as Step~2 in Algorithm~3) performs the computation of supremal quantitatively completable sublanguage, and Step~3 supremal controllable sublanguage, so the iterations of Steps~2---4 generates the following sequence of languages:
\begin{align*}
K &= K^1 = K_1^1 \supseteq NK_1^1 \supseteq ... \supseteq NK_{M-1}^1 \supseteq NK_{M}^1 \\
  &\supseteq K^2 \supseteq ...
\end{align*}
Since Algorithms~4 is finitely convergent, so is the above sequence of languages. When the sequence converges, i.e. in Step~4 $K^{j+1} = K^j$ holds for some $j$,  $K^{j+1}$ is the supremal controllable and quantitatively completable sublanguage of $K$ wrt. $\{(Q_{m,i}, N_i)| i \in \mathcal{I}\}$. This proves that $CK = K^{j+1} = \sup\mathcal{CQC}(K, $ $\{(Q_{m,i}, N_i)|  i \in \mathcal{I}\})$. \hfill $\blacksquare$

By the above proof, the complexity of one complete iteration over all $i \in \mathcal{I}$ in Step 2 is $O(|Q|\cdot |X| \cdot \prod_{i=1}^{|\mathcal{I}|}{N_i})$.
Since Algorithm SC does not increase the state/transition number of ${NK_{M}^j}$ ($M= |\mathcal{I}|$), the complexity of each iteration including Steps 2 and 3 is again $O(|Q|\cdot |X| \cdot \prod_{i=1}^{M}{N_i})$.
Finally since there can be at most $|Q|\cdot |X| \cdot |\Sigma| \cdot \prod_{i=1}^{|\mathcal{I}|}{N_i}$
iterations, the overall time complexity of Algorithm~4 is $O\big((|Q|\cdot |X| \cdot |\Sigma| \cdot \prod_{i=1}^{|\mathcal{I}|}{N_i})^2\big)$. This complexity is the same as that of Algorithm~3.

}

\begin{remark}Another approach of computing
the supremal sublanguage $\sup\mathcal{CQC}(K, \{(Q_{m,i}, N_i)| i \in \mathcal{I}\})$
is to first compute ${\bf K}_i'$ by applying Algorithm 1 with inputs $K_i$ and $N_i$ for each $i \in \mathcal{I}$,
then compute the product automaton
\[{\bf K}' = {\bf SUP} ~|| ~(\mathop{||}_{i\in \mathcal{I}}{\bf K}_i')\]
which represents the supremal quantitatively completable sublanguage,
and finally input ${\bf K}'$ to the algorithm in \cite{QueirozEt:2005} of computing the supremal controllable
strongly nonblocking sublanguage.
According to Algorithm 1, ${\bf K}'$ has at most $|Q|\cdot |X| \cdot \prod_{i=1}^{M}{N_i}$ states, thus this alternative algorithm
has the same with complexity with Algorithm 3.
\end{remark}

\begin{remark}In practice, each bound $N_i$ is specified according to the corresponding task. For an urgent task, the bound could be set small; if not urgent, $N_i$ may be set as a large number. Our proposed solution is
general in handling arbitrary (positive) values set for $N_i$; therefore our solution may be
applicable to a wide range of applications (a few such applications are mentioned in the second paragraph
of Introduction). Our developed algorithms have complexities linear in $N_i$, so are amenable
in handling large bounds. Experimental results will be targeted in our future work, as they are beyond the scope of this brief paper.
\end{remark}

The following example demonstrates Algorithm~4 on synthesizing supervisors satisfying
both controllability and quantitative completability.

\begin{example} [Continuing Example 1]
Consider plant generator ${\bf G}$ and nonblocking supervisor ${\bf SUP}$ displayed in Fig.~\ref{fig:GSUP}.
Input ${\bf G}$, $K = L_m({\bf SUP})$,  {$\mathcal{Q}_{\bf G} = \{Q_{m,1} = \{1,2\}, Q_{m,2} = \{0\}\}$ and $\{N_1 = 3,$ $ N_2 = 5\}$ to Algorithm~\ref{algm:supcsnc}.}

In Step~1, $K^1 := K$.
Then Step~2 generates a language $NK_2^1$ (represented by the generator ${\bf SUP}_2'$ as displayed in Fig.~\ref{fig:SUP2'}); it is
verified that $NK_2^1$ is quantitatively completable wrt. $\{(Q_{m,1} = \{1,2\}, N_1 = 3), (Q_{m,2} = \{0\}, N_2 = 5)\}$. However, it is not controllable, because uncontrollable event 12 is disabled at states 13 (of ${\bf SUP}_2'$).
Next Step~3 generates the supremal controllable sublanguage $K^2$ represented by generator ${\bf NCSUP}_{2}'$ as displayed Fig.~\ref{fig:NCSUP2'}.
Then in Step~4, since $K^2 \neq K^1$, we repeat Steps~2 and 3 and obtain $K^3$. It is verified that $K^3 = K^2$ and thus
Algorithm~\ref{algm:supcsnc} terminates and outputs the language $CK = K^3$ represented by the generator ${\bf QCSUP}$ as displayed Fig.~\ref{fig:NCSUP2'}
(since $K^3 = K^2$, ${\bf QCSUP}$ has the same structure with ${\bf NCSUP}_2'$).
It is verified that $K^3$ is both controllable and quantitatively completable wrt. $\{(Q_{m,1} = \{1,2\}, N_1 = 3), (Q_{m,2} = \{0\}, N_2 = 5)\}$,
and thus according to Theorem~\ref{thm:snsct}, ${\bf QCSUP}$ may be used as a quantitatively nonblocking supervisor.

\begin{figure}[!t]
\centering
    \includegraphics[scale = 0.34]{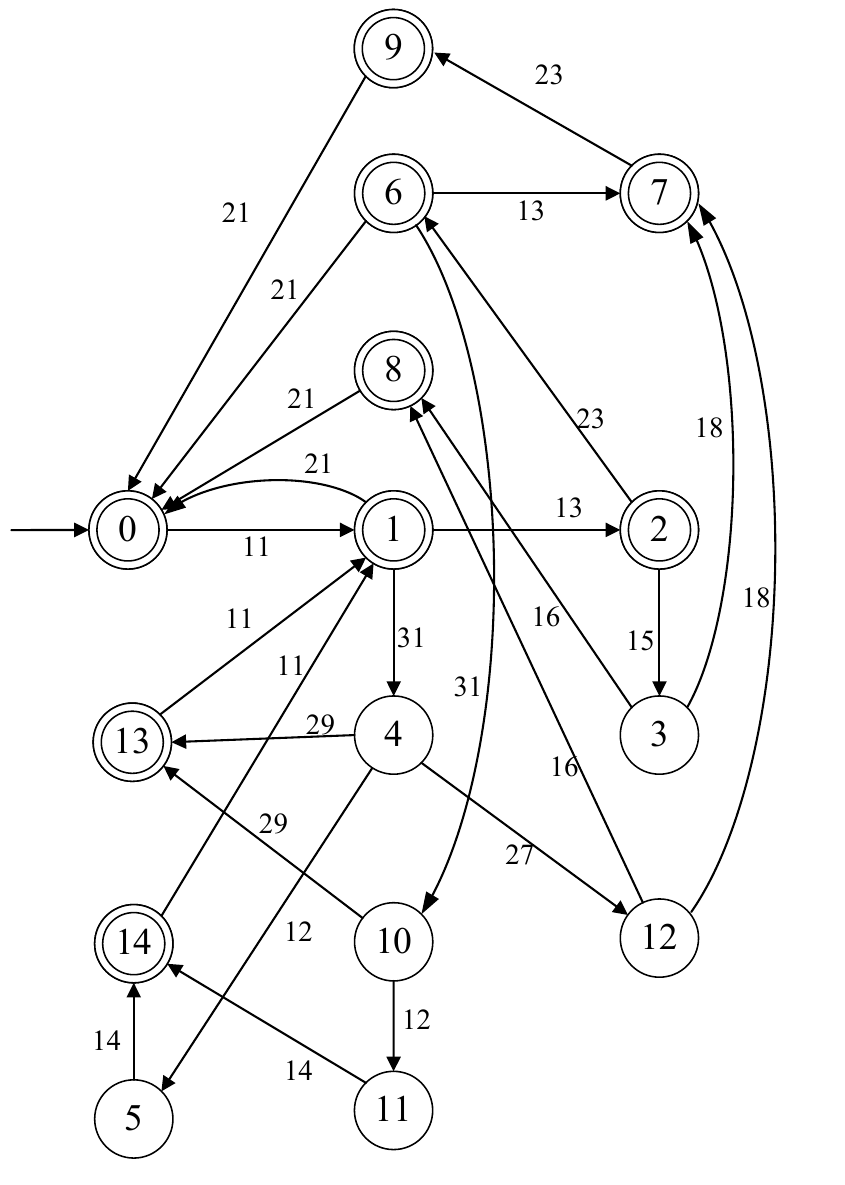}
\caption{Transition graph of generator ${\bf NCSUP}_2'$ and supervisor ${\bf QCSUP}$  (they have identical transition structure)} \label{fig:NCSUP2'}
\vspace{-0.2cm}
\end{figure}

This supervisor ${\bf QCSUP}$ is used to make the autonomous vehicle provide timely services in Example 1.
The control logics of ${\bf QCSUP}$ are as follows: (1) never move to zone~5 when in zone~0; (ii) never move to zone~4 when in zone~3; (iii)
if the vehicle is in zone 1, it is safe to move to zone 2 and zone 4 if it has just returned from zone 0 (i.e. finished self-charging);
and (iv) if the vehicle has moved to zone 3, it should return (either by moving through zone 1, or moving though zone 2 and zone 1) for self-charging before the next round of service.

These logics guarantee that the two requirements ((i) and (ii) in Example~1 of Section 2.2) on the vehicle are satisfied.
First, every package sent to customers can be delivered by the vehicle to one of the two service areas (zone~1 or 2) within three steps; and whenever a customer calls for package collection, the vehicle can reach either zone~1 or 2 within three steps no matter where the vehicle is and no matter which paths (permitted by the supervisor $\bf QCSUP$) the vehicle follows.
{Second, no matter where the vehicle is, it can return to zone 0 for self-charging within five steps no matter which paths the vehicle follows.}
 ~\hfill $\diamond$

\end{example}








\section{Conclusion and Future Work}
{
In this paper, we have introduced a new concept of quantitative nonblockingness,
which requires that every task (each represented by a subset of marker states) must be completed in
prescribed numbers (one for each task) of steps. Moreover, we have formulated a new quantitatively nonblocking supervisory control problem,
characterized its solution in terms of quantitative language completability, and developed algorithms to compute the optimal solution.
}

In this paper the bounds on task completion are specified in terms of the number of transition steps. In practice, the bounds
may also need to be described by the number of time units or even real times. Thus in future work, we are interested in extending the concepts of quantitative nonblockingness to the nonblocking supervisory control framework of timed discrete-event systems \cite{BraWon94, Gouin:1999}.

\small
\bibliographystyle{plain}        
\bibliography{SCDES_Ref}           




\end{document}